\documentclass[12pt,aps]{revtex4}

\usepackage{subfigure}
\usepackage{epsfig}
\usepackage{amssymb}
\usepackage{amsmath}

\newcommand{\chibar}{\overline{\chi}}

\newcommand{\qc}{\bra\,\overline{q}q\,\ket}
\newcommand{\ga}{g_{{\cal A}}}
\newcommand{\gc}{\bra\frac{\alpha_s}{\pi}G^2\ket}
\newcommand{\bra}{{\langle}}
\newcommand{\ket}{{\rangle}}
\newcommand{\bea}{\begin{eqnarray}}
\newcommand{\eea}{\end{eqnarray}}

\begin{document}

\title{Form factors for semileptonic D decays }
\author{Teresa Palmer and Jan O. Eeg}
\affiliation{Department of Physics, University of Oslo,
P.O.Box 1048 Blindern, N-0316 Oslo, Norway}


\vspace{0.5 cm}

\begin{abstract}

We study the form factors for semileptonic decays of $D$-mesons.
 That is, we consider the matrix element of the weak
left-handed quark current for the transitions $D \rightarrow P$ and
$D \rightarrow V$,  where $P$ and $V$ are light pseudoscalar or vector mesons,
respectively.
Our motivation to perform the present study of these form factors are
 future calculations of  non-leptonic decay amplitudes.

We consider the form factors within a class of chiral
 quark models. Especially, we study how  the
Large Energy Effective Theory (LEET) limit works for $D$-meson decays.
Compared to previous work we also introduce light vector mesons
$V = \rho, K^*,...$ within chiral quark models.
In order to determine some of the parameters in our model, 
 we use existing data and results based on some other
methods like 
 lattice calculations,
light-cone sum rules, and heavy-light chiral perturbation theory.
We also obtain some predictions within our framework.

\end{abstract}

\maketitle

\vspace{1cm}


\newpage
\section{Introduction}

In the present paper we study the form factors for semileptonic decays
of $D$-mesons.
Knowledge of the semi-leptonic form factors
is of course necessary to calculate factorizable contributions to the
 non-leptonic decays of mesons. Further, knowledge about these form factors 
might  determine or at least restrict 
 some parameters of our models and thereby 
indirectly be of importance for our (model dependent)   calculations
for non-leptonic decays. 
We are of course aware of the technical challenges  
when calculating
 non-leptonic decays of $D$-mesons \cite{Ryd:2009uf},
  and we will come back to this
 in a future publication.

The $D \rightarrow P$ and
 $D \rightarrow V$ form factors have been calculated by various methods.
The various  frameworks have their strength in different regions
of the momentum transfer  $q$ squared,- from $q^2$ near zero for Light-cone
 sum rules(LCSR) \cite{Ball:1998tj,Ball:1998kk,Khodjamirian:2000ds,Wang:2002zba,Ball:2006yd,DeFazio:2007hw}
to $q^2 = q_{max}^2$ for Heavy-Light Chiral Perturbation
Theory (HL$\chi$PT) \cite{Bardeen:1993ae,Casalbuoni:1996pg,Deandrea:1998uz}.
In  the region  $q^2 \to 0$  where the momentum of the
 outgoing meson is high, one might study form factors within
 Large Energy Effective
 Theory (LEET) \cite{Charles:1998dr}. This effective theory was
later further developed into
Soft Collinear Effective Theory (SCET) \cite{Bauer:2000yr}.

In the region of large momentum transfer ($q^2 \to q_{max}^2$), lattice
QCD can be used
\cite{Flynn:1997ca,Abada:2002ie,Aubin:2004ej,Bernard:2009ke}.
Form factors have been calculated
\cite{Casalbuoni:1996pg,Bajc:1997ey,Fajfer:2004mv,Fajfer:2005ug}
 within 
HL$\chi$PT, which is based on Heavy Quark Effective
 Theory (HQEFT).
Calculations within HL$\chi$PT have also been supplemented
\cite{Hiorth:2002pp} by
calculations within  the Heavy-Light Chiral Quark Model (HL$\chi$QM)
 \cite{Bardeen:1993ae,Ebert:1994tv,Deandrea:1998uz,Hiorth:2002pp,Eeg:2004ik}.
Within the heavy quark symmetry there are corrections of the order
${\cal O}(1/m_c)$ which will be larger in the $D$ sector than in the $B$ sector.
In any case the form factors are  influenced
by nearby meson poles.

Our intention is to find how well 
chiral quark models describe the form factors.
Namely, in the next step we want to calculate nonfactorizable contributions
 to non-leptonic decays of $D$-mesons.
 Then we ought to know how well the chiral quark
 models work in various energy regions, and specifically we need to know the
various  form factors within LEET.
Some form factors are relatively well known.
But some of them might need additional model dependent study beyond
 leading order. Therefore these models will be shortly presented.
Compared to previous work we will in this paper  also include light vectors
$V=(\rho,\omega,K^*$).

\section{Decomposition of Semileptonic form factors}
The $H \to P$ current $J_V^\mu (H \rightarrow P)$ is a vector current
 that depends on the involved momenta  $p_H$ and   $p$.
  This current can be decomposed
into two form factors.
We will consider two commonly used  decompositions
\begin{equation}
J_V^\mu (H \rightarrow P) \; = \;
F_+(q^2) \;
(p_H + p)^\mu
+  F_-(q^2) \;
(p_H -  p)^\mu
 \; \, ,
\label{Fpm}
\end{equation}
or
\begin{equation}
J_V^\mu (H \rightarrow P) \; = \;
F_1(q^2)\left[(p_H + p)^\mu  -\frac{(M_H^2 - m_P^2) }{q^2} q^\mu\right]
+\frac{M_H^2-m_P^2}{q^2} F_0(q^2)q^\mu \; \, ,
\label{F10}
\end{equation}
where $q = p_H - p$ is the momentum transfer.
For decay to  leptons $l$,  $q^\mu  L_\mu$
 ( where $L_\mu$ is the  lepton current)  is
$\sim m_l$ and the amplitude
is dominated by $F_1(q^2)$.
The relations between the form factors in (\ref{Fpm}) and (\ref{F10}) are
\begin{equation}
F_1 \; = \; F_+ \qquad ; \quad F_0 \; =
\; F_+ \, + \; \frac{q^2}{M_H^2-m_P^2} \,F_-  \; \, .
\label{FsRel}
\end{equation}

\begin{figure}[t]
\begin{center}
   \epsfig{file=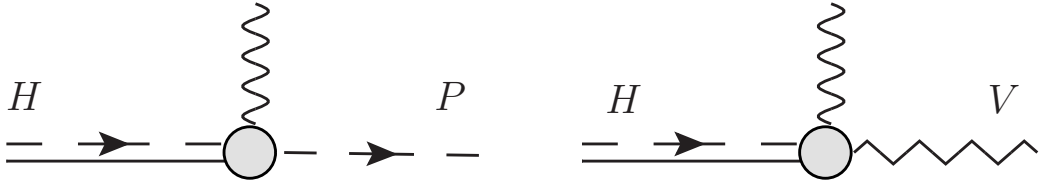,width=15cm}
\caption{
 Diagrams for $H \rightarrow P$ and $H \rightarrow V$ transitions 
at mesonic level.
The  vertical line denotes a virtual  electroweak boson ($W, Z, \gamma$).}
\label{fig:HtoP}
\end{center}
\end{figure}

The semileptonic decays of type $H \to V$, where 
$V= (\rho, K^*, \omega,\phi)$, with mass $m_V$,
 can proceed through both vector and axial currents.
  These can be
decomposed into (in total) four form factors. The vector current
depends on only one form factor $V(q^2)$, and is given by
\begin{equation}
J_V^\mu(H \rightarrow V) \, = \,
\left<V(p,\epsilon)|\bar{q} \gamma^\mu Q|H(p_H) \right>
= \frac{2 V(q^2)}{M_H + m_V} \epsilon^{\mu\nu\rho\sigma}
\epsilon^*_\nu  p_\rho {(p_H)}_\sigma \; \, ,
\end{equation}
while the axial current includes 3 form factors $A_0, A_1, A_2$:
\begin{eqnarray}
J_A^\mu(H \rightarrow V) \, = \,
 \left<V(p,\epsilon)|\bar{q} \gamma^\mu\gamma_5 Q|H\right>
= (M_H+m_V) \left(\epsilon^{*\mu}
-\frac{(  \epsilon^* \cdot q)}{q^2} q^\mu\right) \, A_1(q^2)  \nonumber \\
 - \, \left((p + p_H)^\mu - \frac{M_H^2 - m_V^2}{q^2} q^\mu\right)
\frac{ ( \epsilon^* \cdot q)}{M_H + m_V} A_2(q^2)
+ \, \frac{2 m_V (\epsilon^* \cdot q)}{q^2}  q^\mu A_0(q^2) \; \, .
\end{eqnarray}

For the light leptons ($ l = \mu$, e) the amplitudes for
 $D \rightarrow V \, l \, \nu$
are  dominated by
the form factors
$V(q^2)$, $A_1(q^2)$, and $A_2(q^2)$.
The vector form factor $V(q^2)$ is dominated by vector resonances,
while the $A_1(q^2)$ and $A_2(q^2)$ are dominated
by axial resonances, and the $A_0(q^2)$ form factor is dominated by the
pseudoscalar resonances.

Be\'cirevi\'c and Kaidalov \cite{Becirevic:1999kt} proposed a double
pole form for the $F_+(q^2)$ function.  This includes the pole at
$H^*$ for the first pole and a term
that includes  contributions for higher mass resonances in an
effective pole. The form factors,  $F = F_{+}$, $V$, $A_0$ etc.
 can be written in the generic form:
\begin{equation}
 F(q^2) = \frac{F(0)}{[1-\frac{q^2}{m_{pole}^2}][1-\frac{\alpha \, q^2}{m_{pole}^2}]}
\; ,
\end{equation}
where the parameter $\alpha $ parametrizes the contribution  of the higher
mass resonances into an effective pole.

\section{Asymptotic behavior of  form factors}

HQET and LEET give constraints on the structure of the form factors.
 From HQET one  can
estimate the behavior of the form factors in the limit of zero recoil
(see  \cite{Hiorth:2002pp} and  and references therein): 
\begin{equation}
F_+ \sim \sqrt{M_H}  \qquad ; \qquad F_- \sim \frac{1}{\sqrt{M_H}} \; \, .
\end{equation}

The form factors in the LEET limit, with $p_H^\mu = M_H \, v^\mu$ 
and $p = E \, n^\mu$,
can be parametrized as \cite{Charles:1998dr}:
\begin{equation}
\langle P|\bar{q} \gamma^\mu Q_v|H \rangle \,  =  \,
2E \left( \zeta \, n^\mu + \zeta_1 \, v^\mu \right) \; .
\label{AssHtoP}
\end{equation}

The four vectors $v,n$ are given by $v=(1;\vec{0})$ and $n=(1;0,0,1)$
in the rest frame of the decaying heavy meson.
Here the $\zeta$ should scale with energy $E$ as:
\begin{equation}
\zeta \, \equiv \, \zeta(M_H,E)  \, =
\, C \, \frac{\sqrt{M_H}}{E^2} \quad ,
\; \,  C \sim (\Lambda_\textrm{QCD})^{3/2} \qquad , \quad
\frac{\zeta_1}{\zeta} \sim \frac{1}{E} \; \; .
\label{eq:charlesEq}
\end{equation}
In the limit $M_H \to \infty$ and $E \to \infty$, the ratio
$\zeta_1/\zeta \to 0$.
LEET can be used to estimate  form factors at large recoil, where the momentum
carried by the electroweak bosons ($W, Z, \gamma$)
is at a minimum,
$q^2 \to 0$.
Using (\ref{eq:charlesEq})
for small $q^2$ i.e. for  $E \simeq M_H/2$, one obtains the behavior
\begin{equation}
 F_+ \sim F_0 \sim \frac{1}{M_H^{3/2}}  \; \, .
\end{equation}

For transitions $H(0^-) \rightarrow V(1^-)$ one obtains in the LEET limit
($M_H \to \infty$ and $E \to \infty$) for the vector current:
\begin{equation}
\langle V|\bar{q} \gamma^\mu Q_v |H \rangle = 2i E \zeta_\perp \;
 \epsilon^{\mu\nu\rho\sigma} v_\nu n_\rho \epsilon^*_\sigma \; .
\label{HECurrentV}
\end{equation}
Here the form factor $\zeta_\perp$ 
 scales in
the same way as $\zeta$ in (\ref{eq:charlesEq}), but with a different factor
 $C$:
\begin{equation}
\zeta_\perp = C_\perp \frac{\sqrt{M_H}}{E^2} \quad .
\label{zetascalingPerp}
\end{equation}
For the axial current, the corresponding matrix element should have the form
\begin{equation}
\langle V|\bar{q} \gamma^\mu \gamma_5 Q_v |H \rangle \, = \,
2 E \zeta_\perp^{(a)}  \left[\epsilon^{* \mu} \; - \;
 (\epsilon^{*}\cdot  v) \,  n^\mu \right]
 \; + \;  2 m_V \zeta_{||}(\epsilon^* \cdot v) \, n^\mu \; \, .
\label{HECurrentA}
\end{equation}
Here the form factor $\zeta_\perp^{(a)}$ is equal to
 $\zeta_\perp$ to leading order, and $\zeta_\perp^{(a)}$ and $\zeta_{||}$
scale in the same manner as $\zeta_\perp$ and $\zeta$.

We will need the following relations between the various form factors
 and the quantities $\zeta_i$ in the LEET case
\begin{equation}
F_1 \; = \; F_+ \, = \, \zeta \, + \frac{E}{M_H} \zeta_1
\qquad ; \quad
 F_- \, = \, - \zeta \, + \frac{E}{M_H}\zeta_1 \; ,
\label{FormfHP1}
\end{equation}
It should be noted that in \cite{Charles:1998dr} $\zeta_1$ is neglected 
because it is suppressed by $1/E$.

We will also need the following relations between the various form factors
$V,A_0,A_1,A_2$ and the quantities $\zeta_i$ in the LEET case
\cite{Charles:1998dr}:
\begin{equation}
V \; =  \, \left(1 + \frac{m_V}{M_H} \right) \, \zeta_\perp  \quad ; \;
\quad
A_0 \; =  \;  \frac{m_V}{M_H} 
 \, \zeta_\perp^{(a)}\, + \,  \left( 1 \, - \, \frac{m_V^2}{ M_H E} \right)
 \, \zeta_{||} \; \, ,
\nonumber
\label{FormfHV0p}
\end{equation}
\begin{equation}
A_1 \; =  \, \left(\frac{2 E}{M_H + m_V} \right) \, \zeta_\perp^{(a)} \quad ;
\quad A_2 \; =  \;  \left(1 + \frac{m_V}{M_H} \right) \,\left[ \zeta_\perp^{(a)}
 \, - \frac{m_V}{E} \, \zeta_{||} \right] \; ,
\label{FormfHV}
\end{equation}
which should be valid in the  $q^2 \rightarrow 0$ limit.
These form factors are plotted in section \ref{sec:HtoVff}.

\section{Heavy-Light Chiral Perturbation Theory (HL$\chi$PT)}

HL$\chi$PT is based on heavy Quark Effective Field Theory(HQEFT),
where - to lowest (zeroth) order in $m_Q$ the $0^-$ and the $1^-$ are
degenerate and described by the field
and  $H_v$ is the corresponding
heavy $(0^-,1^-)$ meson field:
\begin{equation}
H_v = P_+(v)\left(\gamma\cdot P^* - i\gamma_5 \,
P_5\right) \; ,
\end{equation}
where $P_+(v) = (1 + \gamma \cdot v)/2$ is a projection operator
and $v$ is the velocity of the heavy quark.
Further,
$P^*_\mu$ is  the  $1^-$ field  and $P_5$ the $0^-$ field.
These  mesonic fields enter the Lagrangian of
 HL$\chi$PT:
\begin{equation}
{\cal L}_{\mathrm{HL}\chi\mathrm{PT}}  \, =
 \, - Tr(\bar{H_v} \, i v_\mu \partial^\mu H_v)
\,  + \, Tr(\bar{H_v}^a \, H_v^b \, v_\mu \, \mathcal{V}^\mu_{ba})
 \, - \ga \, Tr(\bar{H_v}^a \, H_v^b \, \gamma_\mu \, \gamma_5
\mathcal{A}^\mu_{ba}) \; \, ,
\label{eq:L_HLchiPT}
\end{equation}
where $a,b$ are SU(3) flavor indices, and  $g_A$ = 0.59 is the axial coupling.
Further,
$\mathcal{V}_\mu$ and $\mathcal{A}_\mu$ are vector and axial vector fields,
 given by
\begin{equation}
 \mathcal{V}_\mu \equiv \frac{i}{2}(\xi^\dagger \partial_\mu\xi +
 \xi\partial_\mu\xi^\dag) \quad , \quad
 \mathcal{A}_\mu \equiv -\frac{i}{2}(\xi^\dagger \partial_\mu\xi -
 \xi\partial_\mu\xi^\dag)  \; \; ,
\label{VAfields}
\end{equation}
where
\begin{eqnarray}
            \xi \,= \; \exp\{i \Pi/f\} \, , \quad
\Pi &=& \left(\begin{array}{ccc}
     \frac{\pi^{0}}{\sqrt{2}} + \frac{\eta}{\sqrt{6}}
     & \pi^{+} & K^{+} \\
     \pi^{-} & -\frac{\pi^{0}}{\sqrt{2}} +
     \frac{\eta}{\sqrt{6}} & K^{0} \\
     K^{-} & \bar{K}^{0} &
     -\frac{2\eta}{\sqrt{6}}\end{array}\right) \; ,
\label{xidef}
\end{eqnarray}
where $\eta \equiv \eta_8$.

Based on the symmetry of HQEFT,
 the bosonized current for decay of a system with one heavy quark
and one light quark ($Q_v \bar{q}$) forming $H_v$ is 
\cite{Colangelo:1994es,Casalbuoni:1996pg}:
\begin{equation}
 \overline{q_L} \,\gamma^\mu\, Q_{v} \;  \longrightarrow \;
    \frac{\alpha_H}{2} Tr\left[\xi^{\dagger} \gamma^\mu
L \,  H_{v} \right]
  \; ,
\label{J(0)}
\end{equation}
where $Q_{v}$ is a heavy quark field, $v$ is its velocity, and
$H_{v}$ is the corresponding heavy meson field.
This bosonization has to be compared with the
 matrix elements defining  the meson decay
constants $f_H$ (where $H=B,D$) are the same when QCD corrections below
 $m_Q$ are neglected (see \cite{Neubert:1993mb,Eeg:2004ik}):
\begin{equation}
\alpha_H =  f_H  \sqrt{M_H} \; \, .
\label{fb}
\end{equation}
Using the double pole parametrization, form factors were calculated in
 \cite{Fajfer:2004mv}:
\begin{equation}
F_+(q_{max}^2) =  \frac{\alpha_H}{2 \sqrt{M_H} f} \, \ga \,
 \frac{M_H}{m_P+\Delta_{H^*}}
+ \frac{\tilde{\alpha}}{2 \sqrt{M_H}f} \, \tilde{g} \,
 \frac{M_H}{m_P + \Delta{H'^*}} \; \; .
\end{equation}
The term with $\tilde{\alpha}$ and $\tilde{g}$ is the contribution from the
higher resonances. (In  \cite{Hiorth:2002pp} the higher resonance term was not included. Instead some non-pole terms were included).
\begin{figure}[t]
\begin{center}
   \epsfig{file=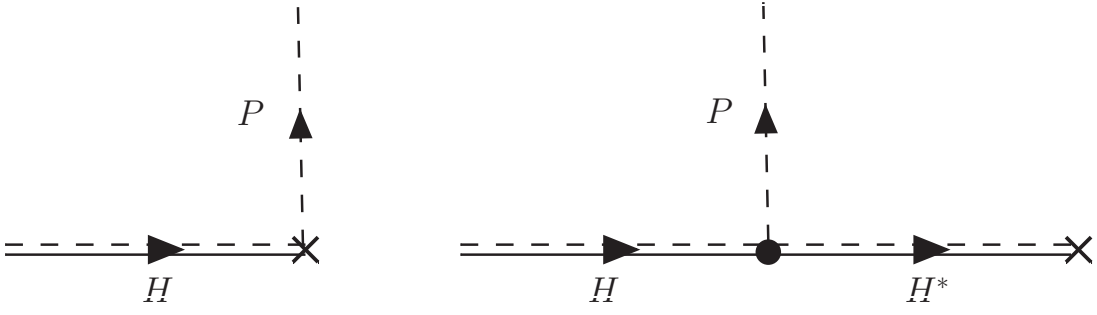,width=15cm}
\caption{Contributions to $F_+$ within HL$\chi$PT.
 The single pole term to the right.}
\label{fig:F+HP}
\end{center}
\end{figure}
One can also include light vectors with an effective coupling to 
heavy mesons, given by
\cite{Fajfer:2005ug}
\begin{equation}
{\cal L}_{\mathrm{HHV}}  \, =
 \, i \frac{g_V}{\sqrt{2}} \,
\lambda \, Tr \left( \bar{H_v} \, H_v \, \sigma_{\mu \nu} \; F_V^{\mu \nu} \right)
\; ,
\label{eq:L_HHV}
\end{equation}
where the coupling $g_V \simeq 5.9$ and  
\begin{equation}
F_V^{\mu \nu} = \partial^\mu V^\nu - \partial^\nu V^\mu \, +
 [V^\mu, V^\nu] \; .
\label{FVtens}
\end{equation}
  This term
will give a dominating  pole term in the $D \rightarrow V$ form
 factor similar to
the one for $D \rightarrow P$ above.
From (\ref{eq:L_HHV}) one obtains \cite{Fajfer:2005ug}:
\begin{equation}
V(q_{max}^2) =  - \, \frac{\alpha_H}{2 \sqrt{M_H} f} \, 
\frac{ g_V \lambda}{\sqrt{2}} \,
 \frac{M_H}{m_V+\Delta_{H^*}}
+ \frac{\tilde{\alpha}}{2 \sqrt{M_H}f} \, \tilde{\lambda} \,
 \frac{M_H}{m_V + \Delta{H'^*}} \; \; ,
\end{equation}
where the second term is coming from higher resonances.
It might also be calculated
 in HL$\chi$QM following closely the calculation for
 $D^* \rightarrow D \gamma$ \cite{Hiorth:2003gd}. 
The coupling
$\tilde{\lambda}$ is  a corresponding term for higher resonances.

\begin{figure}[t]
\begin{center}
   \epsfig{file=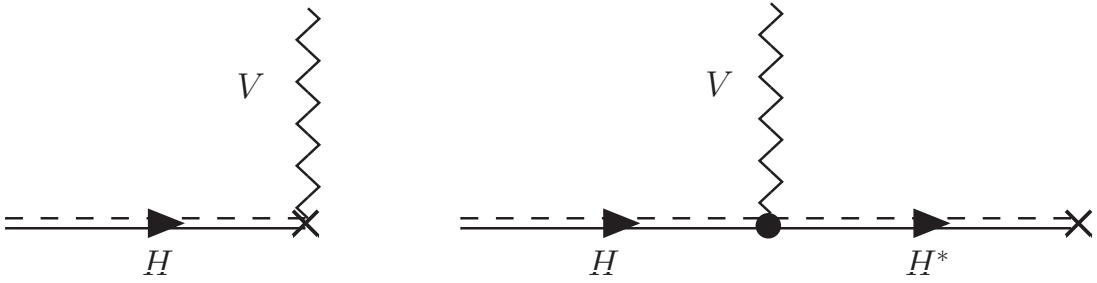,width=15cm}
\caption{Contributions to $H \rightarrow V$ form factors  within HL$\chi$PT.
 The single pole term to the right.}
\label{fig:FHtoV}
\end{center}
\end{figure}

\section{The various Chiral Quark Models}

Calculating the matrix elements of quark currents, we have  used
 chiral quark models.
Within such models one splits the various quark fields into different
categories;  the ordinary soft quark fields
$q$, the soft flavor rotated fields $\chi$ (representing soft
 constituent light quarks),
the heavy quark field $Q_v$ and the light energetic quark field $q_n$.
Moreover, these models might contain ordinary chiral meson fields of $\chi$PT,
as well as energetic light meson fields. They contain light soft pseudoscalar
mesons via ${\cal A}_\mu$ and $\xi$, hard light pseudoscalar mesons in 
an octet filed $M_n$,
the heavy meson fields of $H_v$ of HL$\chi$PT,  and they might contain
 vector meson fields $V^\mu$ with low energy - or vector meson fields
 $V_n^\mu$ with high energy. Here the subscript $n$ in $M_n$ and $V_n^\mu$
refers to the hard momentum $p^\mu = E \, n^\mu$ in 
(\ref{AssHtoP}).  Below we will give a short descriptions
 of the various chiral quark models employed.

\subsection{The $\chi$QM for low energy light quarks}

For the pure light sector, the chiral quark model gives the interactions
between light quarks and light pseudoscalar mesons.
The $\chi$QM Lagrangian
can be written as
 \cite{Manohar:1983md,Bijnens:1985ry,Espriu:1989ff,Eeg:2004ik,Bertolini:1997nf}:
\begin{equation}
{\cal L}_{\chi QM} =   \bar{q}(i\gamma^\mu D_\mu  -  {\cal M}_q) q
  -     m(\bar{q}_R \Sigma^{\dagger} q_L   +    \bar{q}_L \Sigma q_R) \; ,
\label{chqmU}
\end{equation}
where $q$ is the light quark flavor triplet, ${\cal M}_q$ is the current
 mass matrix, and $\Sigma = \xi \cdot \xi$
contains the light pseudoscalar mesons. (The  current mass term
 ${\cal M}_q$ will often be neglected). The covariant derivative
$D_\mu$  contains soft gluons which might
 form gluon condensates within the model. The quantity $m$ is interpreted as
 the constituent light quark mass appearing after the 
spontaneous symmetry breaking
 $SU(3)_L \times SU(3)_R \to SU(3)_V$. 
The Lagrangian  (\ref{chqmU}) can be transformed into a useful version
  in terms of  the flavor rotated fields $\chi_{L,R}$:
\begin{eqnarray}
                 \chi_L =  \xi^\dagger \, q_L \qquad , \quad
                 \chi_R =  \xi  q_R \; \, .
\label{xirot}
\end{eqnarray}
The Lagrangian in (\ref{chqmU}) is then rewritten in the form:
\begin{equation}
 {\cal L}_{\chi QM} = \bar{\chi}\left[ \gamma \cdot (i D + \mathcal{V})
       +\gamma \cdot \mathcal{A} \, \gamma_5 - m \right]  \chi  \; - 
 \chibar \tilde{M_q} \chi \; \, ,
\label{Light-Lagr}
\end{equation}
where the fields $\mathcal{V}$ and $\mathcal{A}$ are given in
equation (\ref{VAfields}), and where the term including the current mass matrix
$ {\cal M}_q$  is given by
\bea
\tilde{M_q} \, = \;  \tilde{M}_q^V  \; + \;   \tilde{M}_q^A \gamma_5  \; ,
\eea
where
\bea
 \label{cmass}
\tilde{M}_q^V \; = \;  \frac{1}{2}(\xi {\cal M}_q \xi  + 
\xi^\dagger {\cal M}_q^\dagger
\xi^\dagger )\quad \text{and} \quad 
\tilde{M}_q^A \; = \;  \frac{1}{2}(\xi {\cal M}_q \xi  
- \xi^\dagger {\cal M}_q^\dagger
\xi^\dagger ) \; .
\eea
This term has to be taken into account  when calculating 
SU(3)-breaking effects.

\subsection{$\chi$QM including light vector mesons (V$\chi$QM)}

The V$\chi$QM adds light vector mesons to the $\chi$QM. The vector
meson fields $V_\mu$ are  given as $\Pi$ in (\ref{xidef})
 with  pseudoscalars $P=(\pi,K,\eta)$ replaced by vectors
 $V=(\rho,\omega,K^*,\phi)$:
\begin{equation}
V_{\mu}=
\left(
\begin{array}{c c c}
\frac{\rho^0}{\sqrt{2}}+ \frac{\omega}{\sqrt{2}} & \rho^+ & K^{*+} \\
\rho^- & -\frac{\rho^0}{\sqrt{2}}+\frac{\omega}{\sqrt{2}} & K^{*0}\\
K^{*-} & \overline{K}^{*0} & - \phi \\
\end{array}
\right) \; .
\end{equation}
These fields are coupled to the light quark fields by the interaction
Lagrangian,
\begin{equation}
{\cal L}_{IV} \, =  \, h_V \, 
\chibar  \gamma^\mu \,  V_{\mu} \, \chi \; . 
\label{chqmV}
\end{equation}
The coupling constant $h_V$ can be determined from  the left-handed
current for \\
$vac\rightarrow V$ we find the $SU(3)$ octet current
\begin{equation}
J_\mu^a(vac \rightarrow V) = \frac{1}{2} m_V f_V Tr \left[ \Lambda^a V_\mu
  \right] \; ,
\label{CurA}
\end{equation}
where the quantity $\Lambda^a$
 is given by $\Lambda^a \, = \xi \lambda^a \xi^\dagger$, and $ \lambda^a$ is
 the relevant SU(3) flavor matrix.
For the currents we obtain
\bea
m_V \, f_V \, = \, \frac{1}{2} h_V \left(- \frac{\qc}{m}+ f_\pi^2 -
\frac{1}{8 m^2} \gc \right) \; ,
\eea
which can be used to determine $h_V$.  
We find, by using $f_\rho \simeq$ 216 MeV, that $h_V \simeq 7$ for standard
values of $m$, $\qc$ and $\gc$ \cite{Hiorth:2002pp,Hiorth:2003gd,Eeg:2004ik}.
\subsection{The Heavy-Light Chiral Quark Model HL$\chi$QM}

The  HL$\chi$QM adds heavy meson and heavy quark fields to the $\chi$QM.
The heavy quark field $Q_v$ is related to the full
field $Q(x)$ in the  following way:
\begin{equation}
Q_v^{(\pm)}(x)=P_{\pm}e^{\mp im_Q v \cdot x} Q(x) \, ,
\label{eq:RHQF}
\end{equation}
where $P_\pm$ are projection operators $P_\pm=(1 \pm \gamma \cdot
 v)/2$. The heavy quark propagator is $S_v(p) = P_+/(v \cdot p)$.
 The Lagrangian for the heavy quarks is:
\begin{equation}
{\mathcal L}_{HQEFT} =   \pm \overline{Q_v^{(\pm)}} \, i v \cdot D \, Q_v^{(\pm)}
 + {\mathcal O}(m_Q^{- 1}) \; ,
\label{LHQEFT}
\end{equation}
where $D_\mu$ is the covariant derivative containing the gluon fields.

To couple the heavy quarks to light pseudoscalar mesons
 there are additional meson-quark
couplings within HL$\chi$QM \cite{Hiorth:2002pp}:
\begin{equation}
{\cal L}_\mathrm{int} = -G_H \left[\bar{\chi}_a \, \bar{H}^a_v \, Q_v
+ \bar{Q}_v \, H^a_v \, \chi_a\right]  \; ,
\label{LintHL}
\end{equation}
where $a$ is a SU(3) flavor index,  $Q_v$ is the reduced heavy quark field
in (\ref{eq:RHQF}).
The quark-meson coupling $G_H$
is  determined within the HL$\chi$QM to be \cite{Hiorth:2002pp}
\begin{equation}
G_H^2 \; = \; \frac{2 m}{f_\pi^2} \, \rho \; \, ,
\label{GHcoupling}
\end{equation}
where $\rho$ is a hadronic quantity of order one \cite{Hiorth:2002pp} .

The V$\chi$QM can be combined
with HL$\chi$QM,  to give a
reasonable description of the weak current for $D$-meson decays
 $D \rightarrow V$
\cite{Fajfer:2005ug}. 
The coupling of $V^\mu$ to heavy mesons is given by 
eq. (\ref{eq:L_HLchiPT}) with
${\cal V}^\mu \rightarrow h_V \, V^\mu$.
In  \cite{Fajfer:2005ug} the factor $\lambda$ is found  to be
$\lambda = - 0.53$ GeV$^{-1}$ . It might also be calculated
 in HL$\chi$QM following closely the calculation for
 $D^* \rightarrow D \gamma$ \cite{Hiorth:2003gd}. 
Using 
the results of  \cite{Hiorth:2003gd}, we obtain
\begin{equation}
\lambda \; = \; - \frac{\sqrt{2} \, h_V \, \beta}{4 g_V}
\end{equation}
which gives $\lambda \simeq  - 0. 4$ GeV$^{-1}$, in agreement with the value 
$\lambda \simeq  - 0. 41$ GeV$^{-1}$ in \cite{Casalbuoni:1996pg}.

For the direct term $J^\mu(H\to V)$ obtained from a quark loop diagram within like in
Fig.~4 has the form
\begin{equation}
J_{tot}^\mu(H_v \rightarrow V) \; =
 \;    \textrm{Tr} \left\{ \xi^\dagger \gamma^\mu L H_v 
\left[ A \, \gamma \cdot V  \, + \, B \, v \cdot V \, \right] \right\} \; ,
\label{eq:JmutotBosV}
\end{equation}
where $A$ and and $B$ are hadronic parameters containing the couplings
$G_H$ and $h_V$, gluon condensates and the  constituent quark mass.
This expression is analogous to eq. (28) in  \cite{Hiorth:2002pp} for the case
$H \rightarrow P$. 
However, the $D \rightarrow V$ form factor will be dominated by the 
pole term in Fig.~3, right, and we will not go further into the detailed
 structure of $A$ and $B$.

\subsection{The Large Energy  Chiral Quark Model  (LE$\chi$QM)}

The LE$\chi$QM adds high energy light mesons and quarks to the
$\chi$QM.
Unfortunately, the combination of the standard version of LEET
\cite{Charles:1999gy}  with $\chi$QM will lead to
 infrared divergent loop integrals  for $n^2=0$.
Therefore, the following formalism is modified and instead  of  $n^2=0$,
 we use $n^2 = \delta^2$, with $\delta = \nu/E$
where $\nu \sim \Lambda_{QCD}$,  such that $\delta \ll 1$.
In the following we derive a
modified LEET \cite{Charles:1998dr} where we  keep $\delta \neq 0$
with $\delta \ll 1$. We call this construction LEET$\delta$ and
 define  the  {\it almost} light -like vectors
\begin{equation}
 n = (1,0,0,+\eta) \qquad ; \quad
 \tilde{n} = (1,0,0,-\eta) \; \, ,
\end{equation}
where
$\eta = \sqrt{1-\delta^2} $. This gives
\begin{eqnarray}
n^\mu + \tilde{n}^\mu = 2v^\mu \; \,  ,\, n^2 = \tilde{n}^2 =
\delta^2 \, , \,
v\cdot n = v\cdot \tilde{n} = 1 \; \; , \, n \cdot  \tilde{n} \, = 2 -
\delta^2 \, .
\end{eqnarray}
The LEET$\delta$ Lagrangian is \cite{Leganger:2010wu}
\begin{eqnarray}
  {\cal L}_{LEET\delta} \, = \,
 \bar{q}_n \left(\frac{\gamma \cdot  \tilde{n} + \delta}{N} \right)
(i n  \cdot D) q_n \; + \; \mathcal{O}(E^{-1}) \; ,
\label{eq:LEETdelta-Lag}
\end{eqnarray}
where $N^2 = 2 n \cdot \tilde{n} $. For
 $\delta \rightarrow  0$ this  is  the first part of the SCET  Lagrangian.
The  quark propagator is
\begin{equation}
S_n(k) \, = \,
                   \frac{\gamma \cdot n}{N(n\cdot k)} \; \; ,
\label{LEETpropd}
\end{equation}
which reduces to the LEET propagator in the  limit $\delta\rightarrow 0$
(which also means $N \rightarrow 2$). For further details we
 refer to \cite{Leganger:2010wu}.

The term  $\mathcal{O}(E^{-1})$  in (\ref{eq:LEETdelta-Lag}) contains a 
term coming from the current mass $m_q$ for the light energetic quark(s).
We have found that further development beyond \cite{Leganger:2010wu} gives the 
SU(3) breaking mass term:
\begin{eqnarray}
  \Delta{\cal L}_{LEET\delta}(m_q) \, = \, \frac{m_q}{E}
 \bar{q}_n \left(i \tilde{n} \cdot D \; - \;
 \frac{m_q}{2}  \gamma \cdot \tilde{n}  \right) q_n  \; .
\label{eq:LEETdelta-Mass}
\end{eqnarray}

For hard light quarks and chiral quarks coupling
to  a hard light meson multiplet field $M$, we extend the ideas of
$\chi$QM and HL$\chi$QM, and assume that the energetic light mesons
couple to light quarks with a derivative coupling to an axial current:
\begin{eqnarray}
{\cal L}_{\mathrm{int}q}
 \; \sim \;
\bar{q} \, \gamma_\mu \gamma_5(i \, \partial^\mu M) \, q \; \, .
\label{Ansatz}
\end{eqnarray}

The outgoing light  energetic mesons are described by  an octet
$3\times 3$ matrix field $M =\exp{(+ i E n \cdot x)} \,  M_n \, $,
where  $M_n$ has
 the same form as $\Pi$ in (\ref{xidef}):
\begin{eqnarray}
M_n &=& \left(\begin{array}{ccc}
     \frac{\pi^{0}_n}{\sqrt{2}} + \frac{\eta_n}{\sqrt{6}}
     & \pi^{+}_n & K^{+}_n \\
     \pi^{-}_n & -\frac{\pi^{0}_n}{\sqrt{2}} +
     \frac{\eta_n}{\sqrt{6}} & K^{0}_n \\
     K^{-}_n & \bar{K}^{0}_n &
     -\frac{2\eta_n}{\sqrt{6}}\end{array}\right) \; \, .
\label{LEMes}
\end{eqnarray}
Here $\pi^{0}_n$, $\pi^{+}_n$, $K^{+}_n$ etc.
are the (reduced) meson fields
 with momentum $\sim E n^\mu$. Furthermore, $q_n$ is related to $M_n$ in the same manner as $Q_v$ is related to $H_v$. 

Combining  (\ref{Ansatz}) with the use of the rotated
 soft quark fields in (\ref{xirot}) and  using
$\partial ^\mu \rightarrow i E \,  n^\mu$
 we  arrive at the LE$\chi$QM interaction Lagrangian \cite{Leganger:2010wu}:
\begin{eqnarray}
{\cal L}_{LE \chi QM}
  \; = \;
         \, G_A \, E
\bar{\chi} \,  (\gamma \cdot n) \, Z_n \, q_n \, + \, h.c.
\; \, ,
\label{eq:LEETdHLciQM}
\end{eqnarray}
where $q_n$ is the reduced field corresponding to an energetic light
 quark having momentum
fraction close to one (see (\ref{eq:LEETdelta-Lag})),
  and $\chi$ represents a soft quark
 (see Eq. (\ref{xirot})). Further,
$G_A$ is an unknown coupling to be determined later by physical requirements.
Further,
\begin{eqnarray}
Z_n         = \xi M_R \, R - \xi^\dag M_L \, L \; \, .
\end{eqnarray}
Here $M_L$ and $M_R$ are both equal to $M_n$, but they have formally
different transformation properties.

\begin{figure}[t]
\begin{center}
   \epsfig{file=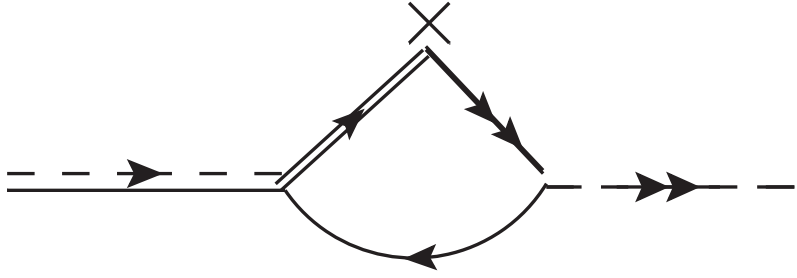,width=11cm}
\caption{Current matrix element in LE$\chi$QM. The double dashed line is 
the (external) heavy meson $H_v$, and the dashed line with two arrows is the 
(external energetic light meson. The internal lines (double for heavy quark, 
single with two arrows is the energetic light quark $q_n$ and with one arrow is
 the soft quark $\chi$.}
\label{fig:DtoPleet}
\end{center}
\end{figure}

Calculating the matrix elements of quark currents within LE$\chi$QM,
we obtain \cite{Leganger:2010wu}:
\begin{equation}
\zeta \; = \; \frac{1}{4}\,  m^2 \, G_H \, G_A \,  F \,
 \sqrt{\frac{M_H}{E}} \; ,
\label{zetaExpr}
\end{equation}
where the quantity $F$ comes from loop integration in Fig.
\ref{fig:DtoPleet} (with soft gluons forming gluon condensates added)
  is \cite{Leganger:2010wu}  :
 \begin{equation}
F \; = \; \frac{N_c }{16 \pi} \; + \;
\frac{3 \, f_\pi^2}{8 m^2 \, \rho} (1- g_A)  \,
 - \,  \frac{(24 - 7 \pi)}{768 \, m^4} \, \gc \; \, ,
\label{FAExpr}
\end{equation}
which is numerically $F \simeq 0.08$.  In Fig.~5, the quantity $F$ is plotted 
 as function of the quark condensate for typical values of the
 constituent quark mass.
 We obtain the following expression for the coupling constant
 \begin{equation}
G_A \; = \;
\frac{4 \zeta}{m^2 \, G_H \, F} \, \sqrt{\frac{E}{M_H}} \; ,
\label{GAExpr}
\end{equation}
where $\zeta$ is numerically known
 \cite{Ball:1998tj,Ball:2004ye,DeFazio:2007hw} to
be $\simeq \, 0.3$
for the transition $B \rightarrow \pi$, but is  larger, say
 $\simeq 0.6$  for
 $D \rightarrow \pi $ \cite{Ball:2006yd}.

 Within our model, the constituent
 light quark mass $m$ is the analogue of $\Lambda_{QCD}$. 
To see the behavior of $G_A$ in terms of the energy $E$ we therefore
 write $C$ in (\ref{eq:charlesEq}) as
$C \, \equiv \, \hat{c} \, m^{\frac{3}{2}}$, and obtain
\begin{equation}
G_A \; = \;
 \left(\frac{4 \hat{c} f_\pi}{m \, F \, \sqrt{2 \rho}} \right)
\;  \frac{1}{E^{\frac{3}{2}}} \; \, ,
\label{GAExpr2}
\end{equation}
which explicitly displays the behavior $G_A \sim E^{-3/2}$. In terms of the
number $N_c$ of colors, $f_\pi \sim \sqrt{N_c}$ and $F \sim N_c$ which gives
the behavior $G_A \sim 1/\sqrt{N_c}$, i.e. the same behavior as for
the coupling $G_H$ in (\ref{LintHL}). We consider $G_A$ to be an auxiliary
quantity. Physical results are expressed in terms of form factors,
and the 
 bosonized current
 can now be written as 
\begin{equation}
J_{tot}^\mu(H_v \rightarrow M_n) =
 - 2 i \zeta \, \sqrt{\frac{E}{M_H}} 
 \; \,   \textrm{Tr} \left\{ 
\gamma^\mu L 
H_v 
\left[\gamma \cdot n \right]
\xi^\dag M_L \right\} \; ,
\label{eq:JmutotBos}
\end{equation}
if $\zeta_1/\zeta \rightarrow 0$ is used.

The LE$\chi$QM can be extended to include energetic vector mesons $V^\mu_n$,
 in analogy
with $M_n$ in (\ref{LEMes}).  In (\ref{Ansatz}) derivative coupling was 
used for coupling of mesons to quarks through an axial vector field.
This is in analogy with light mesons coupling to quarks in (\ref{Light-Lagr}).
As in eq.(\ref{Ansatz}) we will use derivative coupling, here through 
the field $F_V^{\mu \nu}$ in (\ref{FVtens}):
 \begin{eqnarray}
{\cal L}_{LE\chi V} \sim \overline{\chi} \sigma \cdot F_V \, \chi
 \; .
\end{eqnarray}
It was found in  \cite{Leganger:2010wu} that derivative coupling
 gave the best description of the $H \rightarrow P$ current.
Using $V \rightarrow \exp{(i E n \cdot x)} \, V_n$, we obtain
 an interaction (remember that $\partial^\mu V_\mu =0$ implies $n \cdot V_n =0$):
\begin{eqnarray}
{\cal L}_{LE\chi V}
  \; = \;
         \, E \, G_V \,
\bar{\chi} \, \left(\gamma \cdot n  \, \gamma \cdot Z_n \right)
  \, q_n \, + \, h.c.
\; \, ,
\label{eq:LEV}
\end{eqnarray}
where
\begin{eqnarray}
Z_n^\mu         = V_n^\mu \left( \xi \, R  +  \xi^\dag  \, L \right) \;  ,
\end{eqnarray}
 and
\begin{eqnarray}
V^\mu_n &=& \left(\begin{array}{ccc}
     \frac{\rho^{0}_n}{\sqrt{2}} + \frac{\omega_n}{\sqrt{2}}
     & \rho^{+}_n & K^{*+}_n \\
     \rho^{-}_n & -\frac{\rho^{0}_n}{\sqrt{2}} +
     \frac{\omega_n}{\sqrt{2}} & K^{*0}_n \\
     K^{*-}_n & \bar{K}^{*0}_n &
     -\Phi_n \end{array}\right)^\mu \; \, .
\label{LEVes}
\end{eqnarray}
Here $\rho^{0}_n$, $\rho^{+}_n$, $K^{*+}_n$ etc.
are the (reduced) vector meson fields  corresponding to  energetic light
vector mesons
 with momentum $\sim E n^\mu$.
The coupling $G_V$ is determined by the experimental value for the
form factors for 
$B \rightarrow \rho$ (for $B$- decays) or the $D \rightarrow \rho$
(for $D$- decays) at $q^2=0$ obtained either from experiment,
 lattice calculations or eventually LCSR calculations.

In our  case where no  extra  soft  pions are going out,
  we put $\xi \rightarrow 1$,
and for the momentum space $V_n^\mu \rightarrow k_M \sqrt{E}(\epsilon_V^*)^\mu $,
with the isospin factor
 $k_M = 1/\sqrt{2}$ for $\rho^0$ (while $k_M=1$ for charged $\rho$'s).
For the  $D$-meson
with spin-parity  $0^-$ we have
$ H^{(+)}_v
\rightarrow P_+(v)(-i\gamma_5)\sqrt{M_H}$. Using this,
 the involved traces are easily calculated, and we obtain
$J_{tot}^\mu(H_v \rightarrow V_n)$
 for the 
 $H_v \rightarrow V_n$
 transition. 

Using the equations (\ref{eq:charlesEq}), (\ref{FVExpr}), and
(\ref{eq:MtotFTW}), one  obtains \cite{Leganger:2010wu} the relations 
between $G_V$ and $\zeta$. The loop factor will also be $F$ in this case.
The formulae relating $\zeta_\perp$ and $G_V$ will be exactly those
relating $\zeta $ and $G_A$, that is 
\begin{equation}
\zeta_\perp \; = \;
 \frac{1}{4} \, m^2 \,  G_H \, G_V \, F \, \sqrt{\frac{M_H}{E}}
\label{eq:MtotFTW}
\end{equation}
and so on, obtained by the replacements $\zeta \rightarrow \zeta_\perp$
and $G_A \rightarrow G_V$. in (\ref{GAExpr}) and (\ref{GAExpr2}).
Here $\zeta_\perp$ is numerically known  for $B \rightarrow \rho$ to be
$\simeq 0.3$ \cite{Ball:1998kk,DeFazio:2007hw} and for 
$D \rightarrow \rho$ to be $\simeq 0.59$ from CLEO data \cite{CLEO:2011ab}.
 
The bosonized current for the vector case can, for 
 $m/E << 1$ (implying also $\zeta_\perp^{(a)} \rightarrow \zeta_\perp$)
be written as  
\begin{equation}
J_{tot}^\mu(H_v \rightarrow V_n) =
 - 2 i \, \sqrt{\frac{E}{M_H}} 
 \; \,   \textrm{Tr} \left\{ 
\gamma^\mu L 
H_v \, \left(\zeta_\perp \gamma \cdot n \; - \; \frac{m_V}{m} \zeta_{||} \right)
\,  \sigma \cdot F_{n} \,  \xi^\dag \,
\left[\gamma \cdot n \right] \, \right\} \; ,
\label{eq:JmutotVBos}
\end{equation}
where the tensor $F_{n}$ is given by (\ref{FVtens}) with $V_n$ given as in 
(\ref{LEVes}).
We find the following  predictions within our model:
\begin{equation}
\zeta_\perp^{(a)} = \zeta_\perp \, + \frac{m}{E} \zeta_{||} \quad ; \;  
\zeta_{||} \; = \;  \frac{m \, F_{||} }{m_V \, F }\zeta_\perp \quad ; \;
\zeta_1 \;  = \;  \frac{m  \, F_{||}}{E F}\zeta  \qquad ; \quad
\label{zetarel}
\end{equation}
where 
\begin{equation}
F_{||} \; = \; \frac{N_c }{16 \pi} \; + \;
\frac{3 \, f_\pi^2}{8 m^2 \, \rho} (1- g_A)  \, + \,
\frac{f_\pi^2}{2m^2} \ell
 + \,  \frac{1}{48 \, m^4} \, \gc \;
 \left( \frac{7 \pi}{16} - 2  \right) \, ,
\label{FVExpr}
\end{equation}
is a loop function analogous to $F$ in (\ref{FAExpr}). Here the appearance of $\ell = ln(2/\delta)$ is 
due to the 
infrared behavior of some of the loop integrals.  
Numerically one finds $F_{||} \simeq 0.24 \simeq 3 F$.
 
\begin{figure}[t]
\centering

\subfigure{
\includegraphics[width=.48\textwidth]{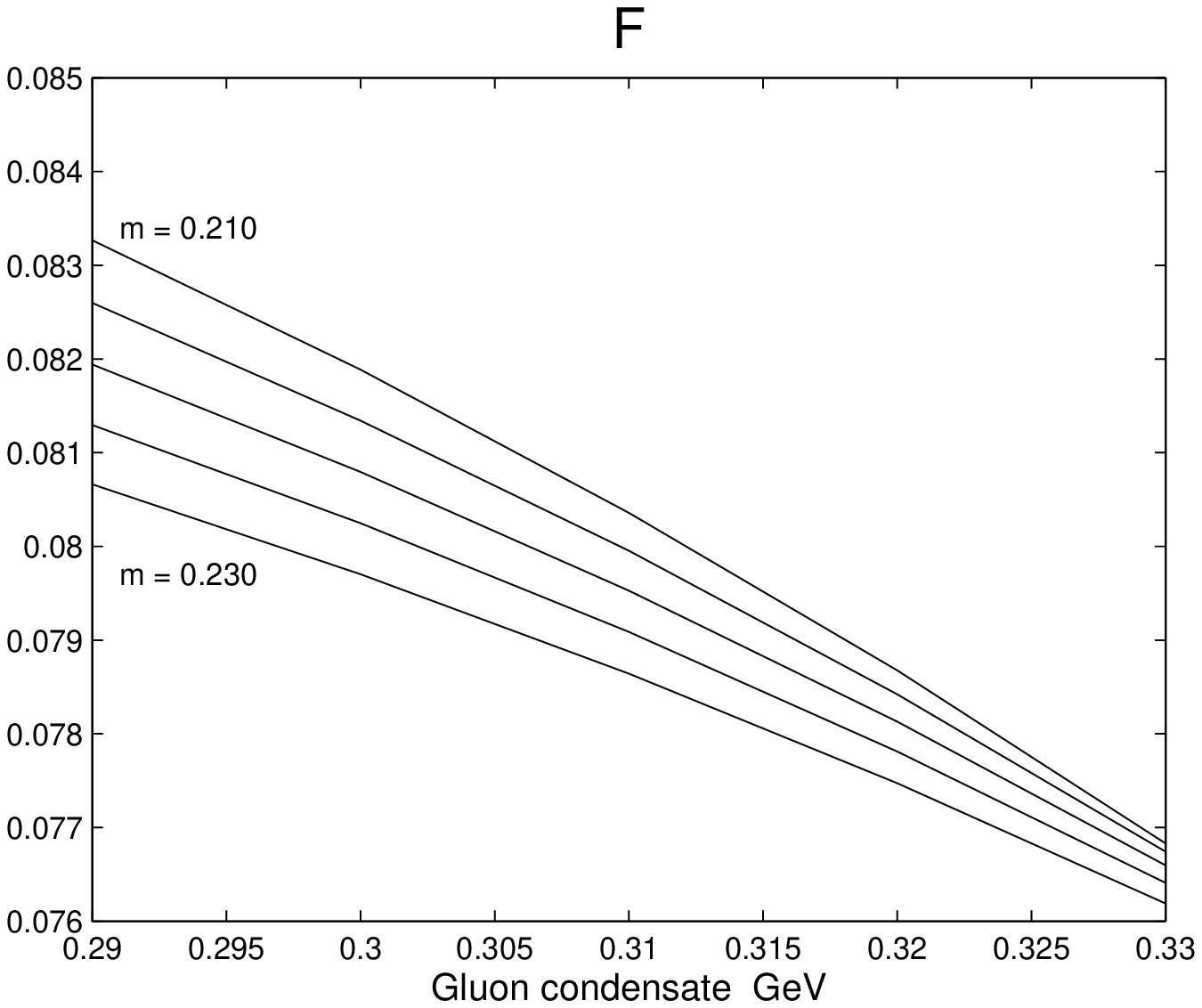}
\label{fig:FA}
}
\subfigure{
\includegraphics[width=.48\textwidth]{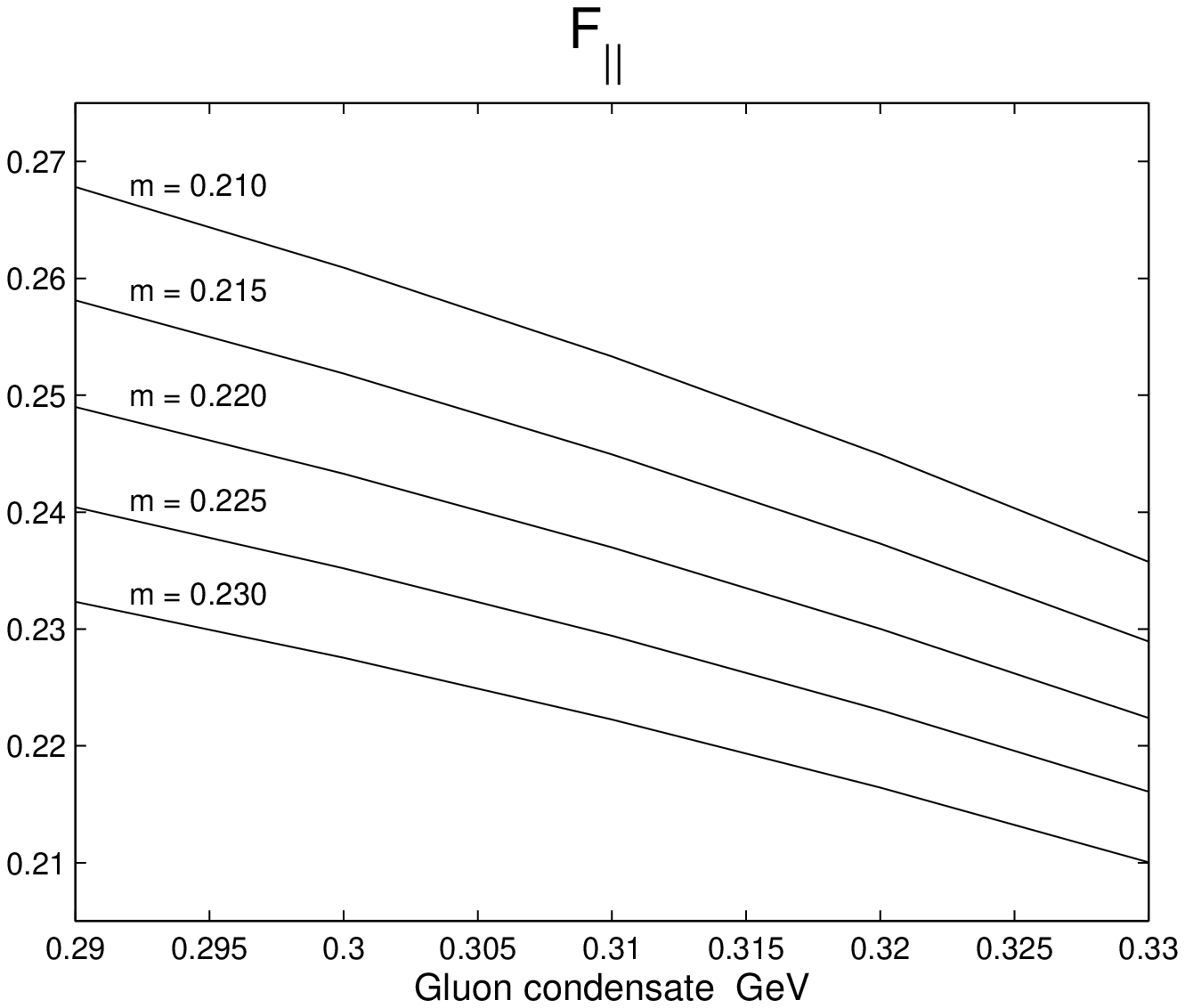}
\label{fig:FV}
}
\label{fig:FaFv}
\caption[Optional caption for list of figures]{$F$ and $F_{||}$ as
 a function of $\left<\frac{\alpha_s}{\pi}G^2\right>^{1/4}$ for values 
of constituent quark mass
from $m$ = 0.210 to $m$= 0.230 GeV}
\end{figure}

The values for $F$ and $F_{||}$ are obtained with the simplified  
LEET propagator in (\ref{LEETpropd}) .
 For the $B \rightarrow D$ 
case an extra $\Delta$
of order 20 MeV was used in the heavy quark propagator 
\cite{Eeg:2009qq}.
A similar assumption (which is closer to 
 the SCET propagator)might be used here,- leading to modified values of
  $F$ and $F_{||}$. 
We observe that although 
 $\zeta_1/\zeta \;  \sim \;  \frac{m}{E}$ as it should, but the numerical  
suppression is not strong because $F_{||} \simeq 3 F$ and $\frac{m}{E}$ 
is not very small for $D$-meson decays.

So far we have  considered the $SU(3)$-limit $m_q \rightarrow 0$.
One  may also calculate $SU(3)$ corrections from the mass correction 
Lagrangian in (\ref{eq:LEETdelta-Mass}), for hard outgoing $s$-quarks.
 We find that the first order term
does not contribute within LE$\chi$QM. The second order term in 
(\ref{eq:LEETdelta-Mass}) contributes and  gives terms suppressed by
$m_s^2/(m \, E)$ compared to terms already calculated. These will therefore 
be discarded in this work. 
For decaying $B_s$ and $D_s$ there will be first order $m_s$ corrections
from the ordinary light sector $\chi$QM, through mass terms 
in (\ref{cmass}). 
 However, these  corrections must 
be considered together with meson loops. Some of  these loops might be 
calculated as in chiral perturbation theory, while others are
formally suppressed, and anyway problematic to handle within our formalism.
  Therefore we do not go further into these details.

\section{Plotting the form factors}
\label{sec:HtoVff}
In this section we
plot form factors for $D \rightarrow P$ and
$D \rightarrow V$ and as function of $q^2$. We have used input from 
experimental data \cite{CLEO:2011ab,pdg:2010},
lattice gauge calculations \cite{Flynn:1997ca,Abada:2002ie},
Light Cone Sum Rules (LCSR) 
\cite{Ball:1998tj,Ball:1998kk,Khodjamirian:2000ds,Wang:2002zba,Ball:2006yd,DeFazio:2007hw} and  
Light Front Quark Model \cite{Verma:2011yw}. 
The plots do not include error bars because that would make them unreadable.
 For LEET, $q^2 =0$ is the reference point, 
 and the shape is  determined
 by a single pole.

For HL$\chi$PT the no-recoil point ($q^2 = (q^2)_{max}$)
 is the reference point for plots, and the form factors might be 
calculated within the framework of HL$\chi$QM.
The plots for  $D \rightarrow P$ with $P= \pi,K,\eta$
will  be different because of the different masses. But we have not explicitly
calculated SU(3)-breaking effects, and (\ref{zetas}) below
 should be valid in the 
SU(3)-limit $m_s \rightarrow 0$.
This means the plots for $D \rightarrow \pi$ and $D \rightarrow \rho$ are 
most relevant for us. Still other plots are included for comparison.
According to our model (see eq. (\ref{eq:LEETdelta-Mass})),
 SU(3) corrections due to hard $s$-quarks
(as in  $D \rightarrow K$ and $D \rightarrow K^*$ transitions) should be small, 
while SU(3) corrections due to soft $s$-quarks (as in decays 
of $D_s$) should be bigger, as pointed out at the end of section V.

From the plots we  extract  approximate  values for the
form factors $F_+(q^2)$, $V(q^2)$, $A_1(q^2)$ and $A_2(q^2)$ at $q^2 =0$.
These are collected in table I and II.

\begin{figure}[!ht]
\centering

\subfigure{
\includegraphics[width=.48\textwidth]{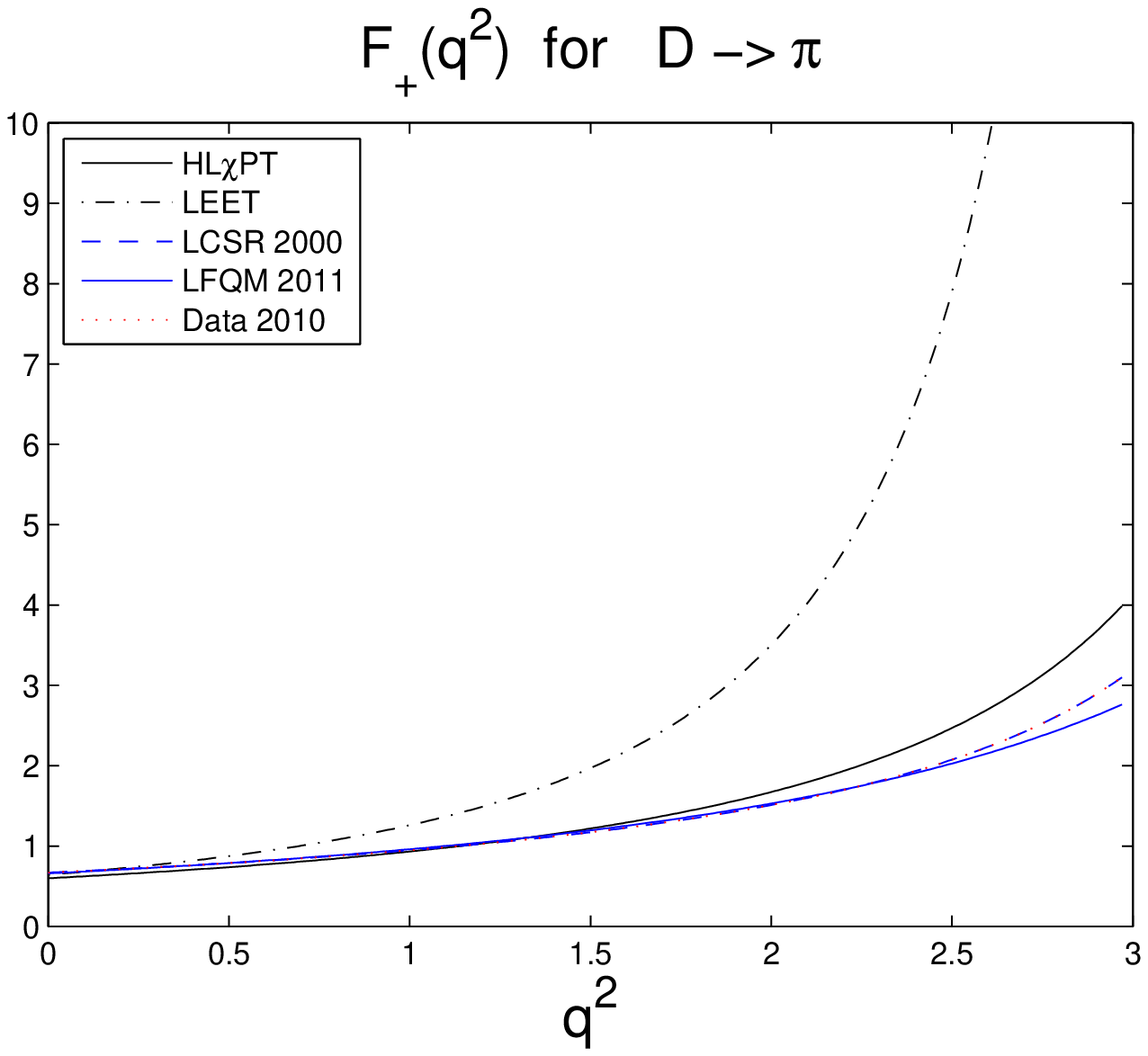}
\label{fig:dpif}
}
\subfigure{
\includegraphics[width=.48\textwidth]{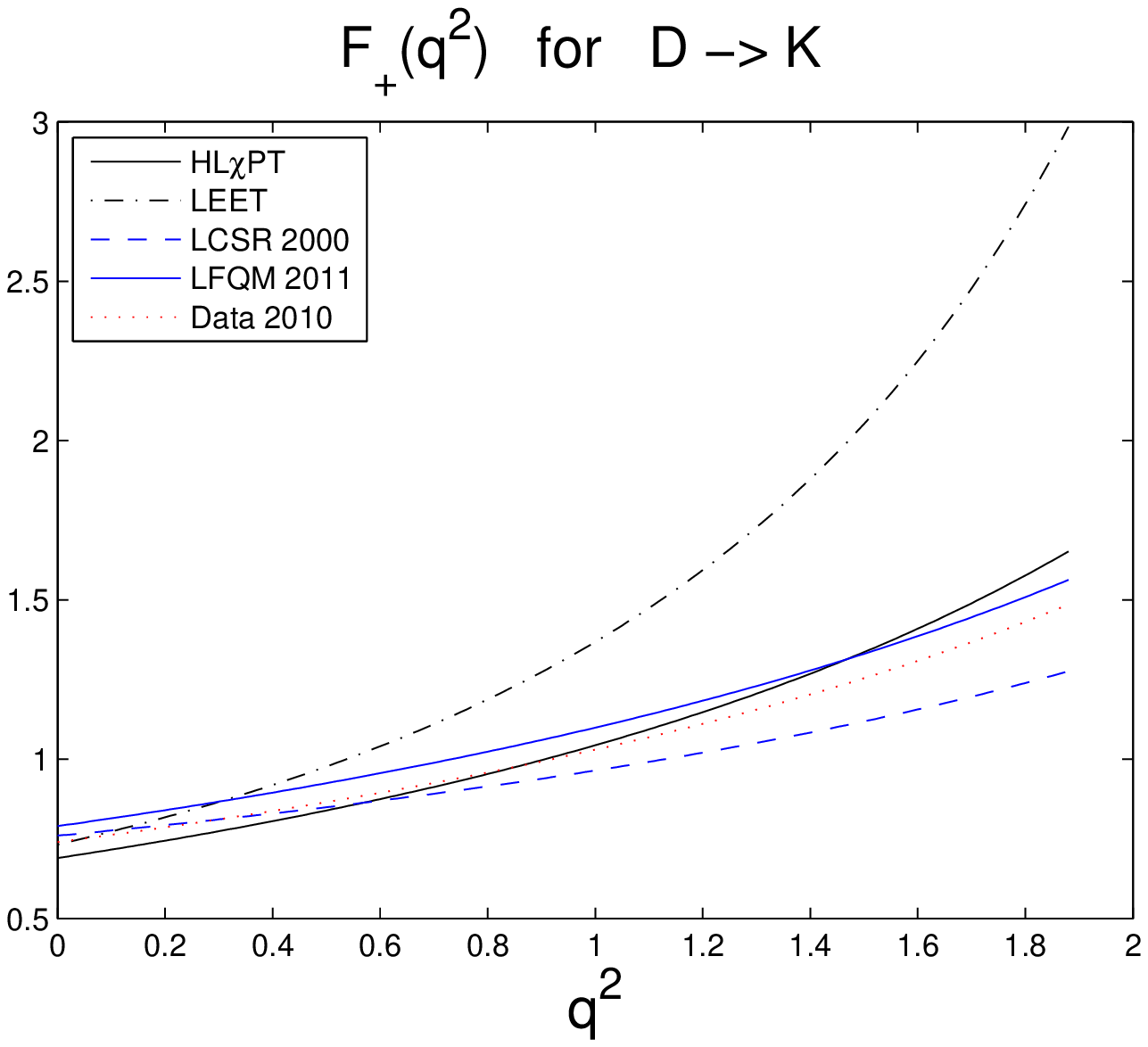}
\label{fig:dkf}
}
\subfigure{
\includegraphics[width=.48\textwidth]{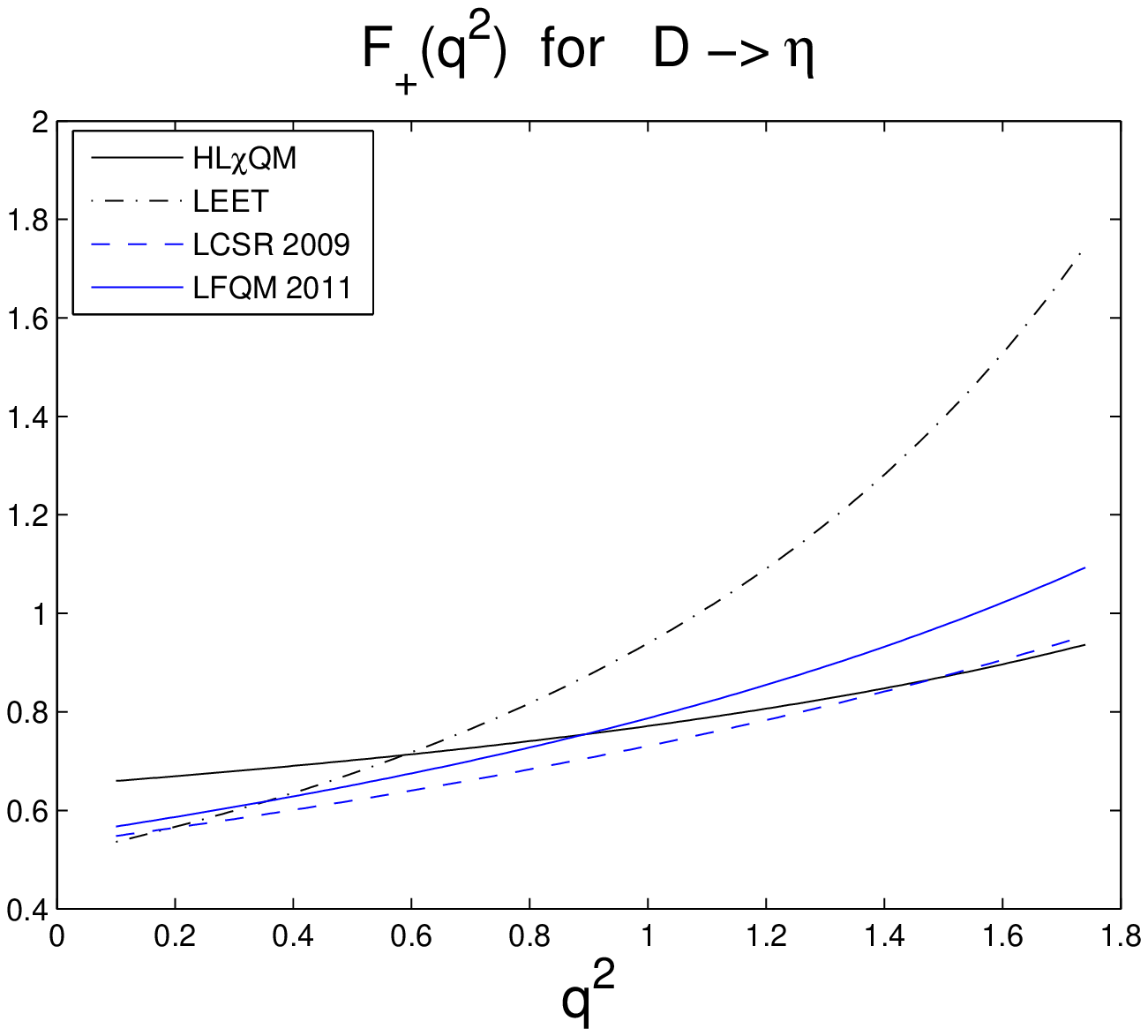}
\label{fig:detaf}
}
\subfigure{
\includegraphics[width=.48\textwidth]{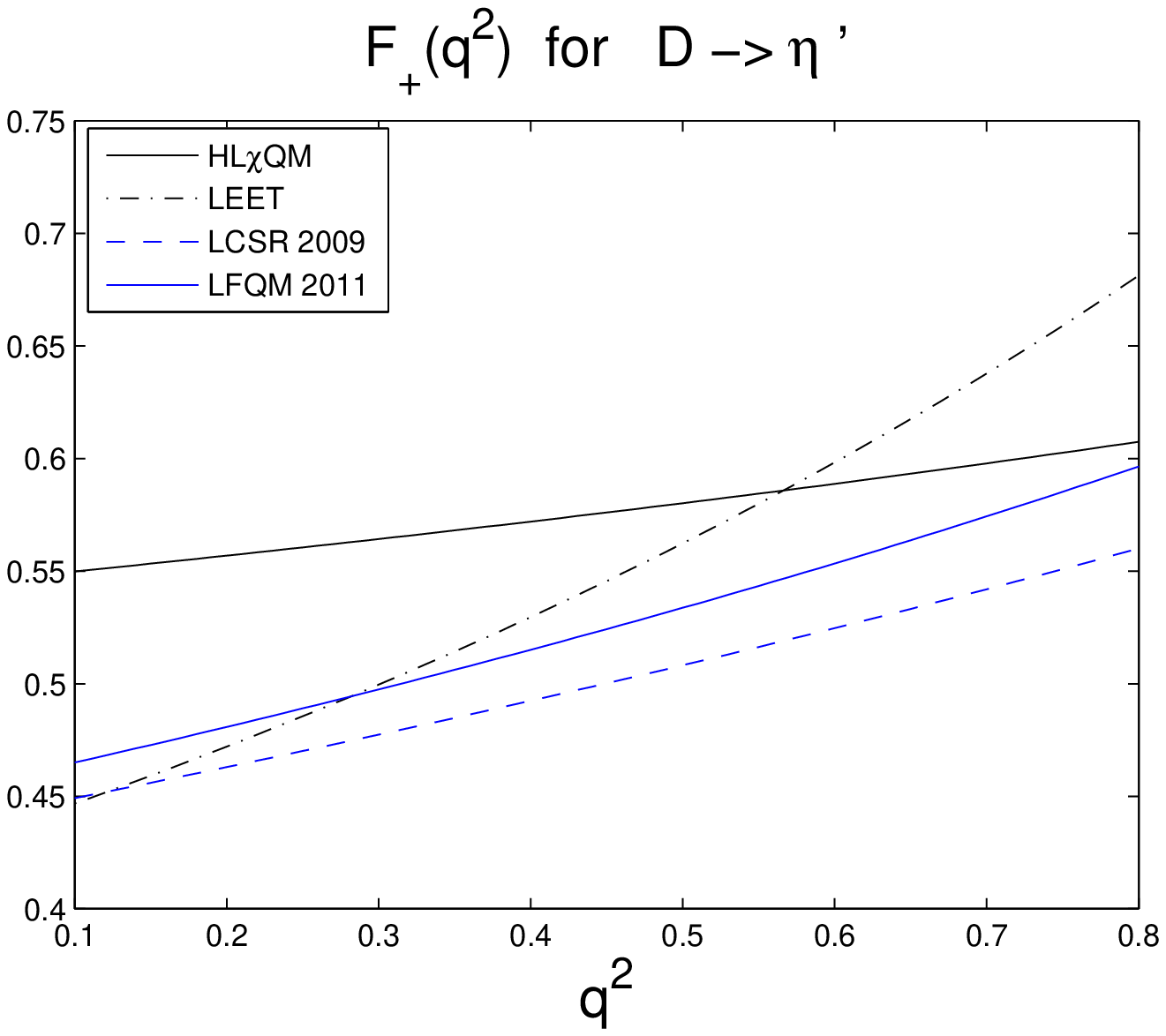}
\label{fig:detaprimef}
}
\label{fig:dpff}
\caption[Optional caption for list of figures]{$D \rightarrow P$ form factors
 comparing frameworks used: 
HL$\chi$PT is from  \cite{Fajfer:2004mv}, LCSR 2000  from 
\cite{Khodjamirian:2000ds},
LCSR 2009 from \cite{Wang:2008zz}, LEET 
from \cite{Charles:1998dr}, LFQM  from \cite{Verma:2011yw}, and 
Data is from \cite{pdg:2010}. }
\end{figure}

\begin{figure}[!ht]
\centering
\subfigure{
\includegraphics[width=.48\textwidth]{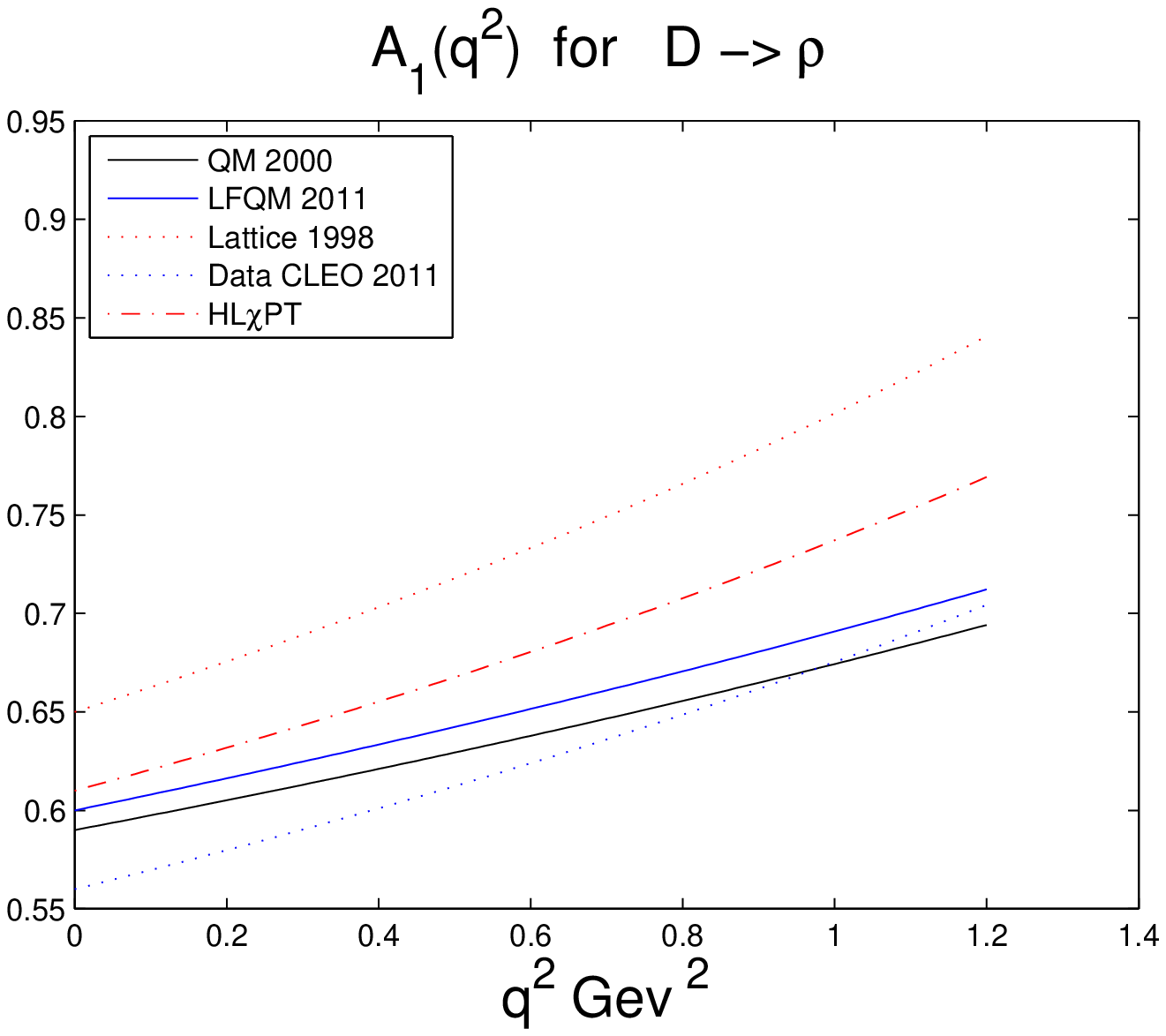}
\label{fig:drhoa1}
}
\subfigure{
\includegraphics[width=.48\textwidth]{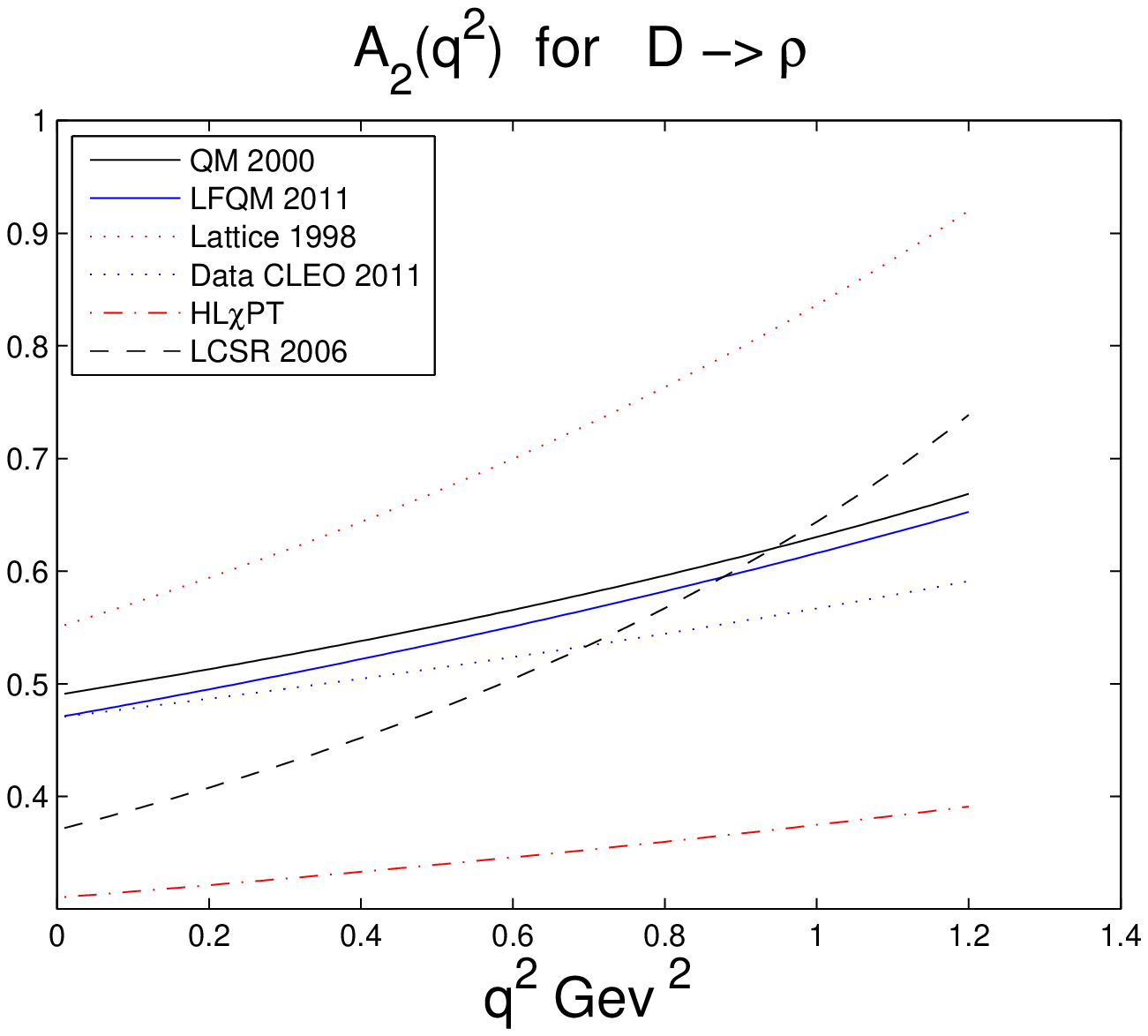}
\label{fig:drhoa2}
}
\subfigure{
\includegraphics[width=.48\textwidth]{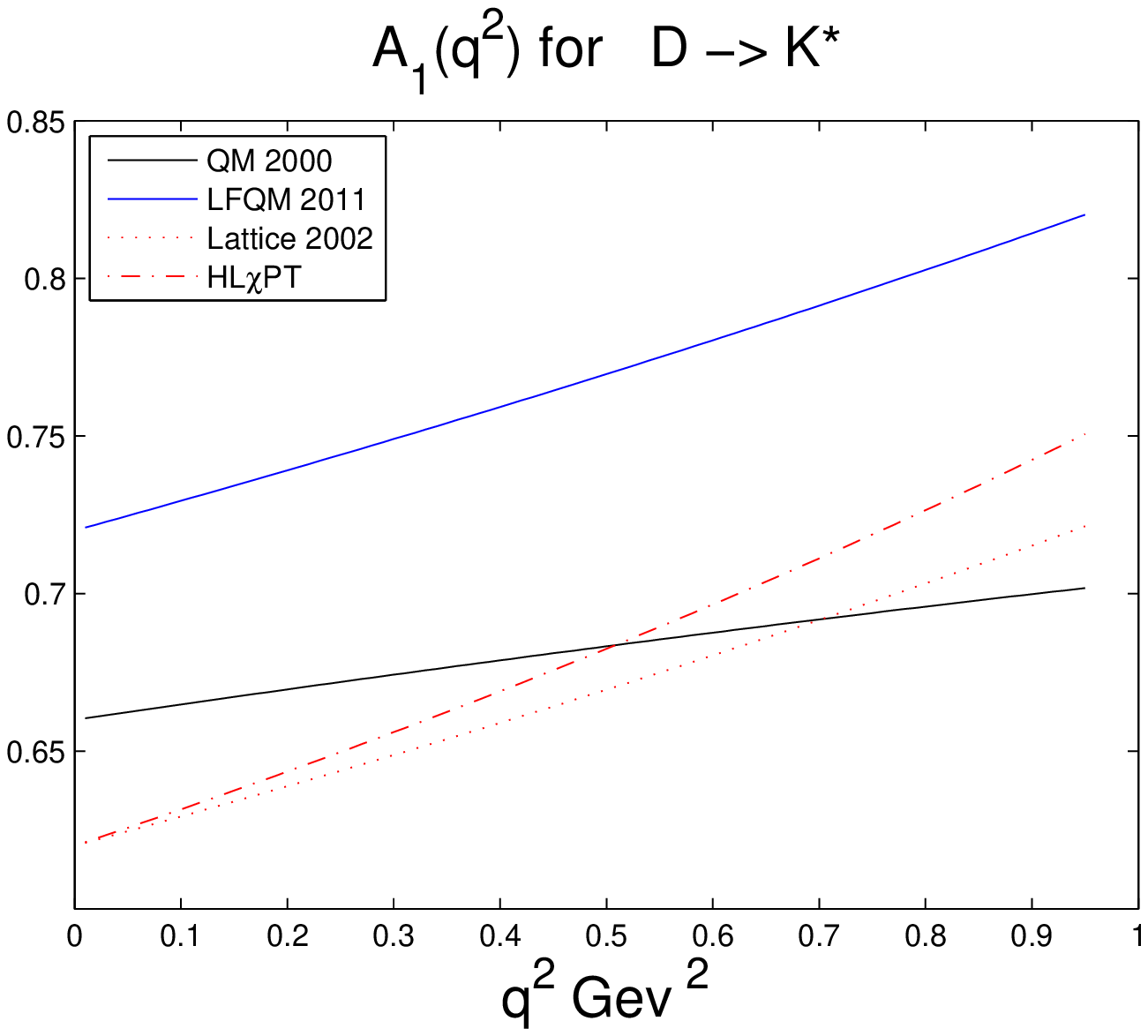}
\label{fig:dkstara1}
}
\subfigure{
\includegraphics[width=.48\textwidth]{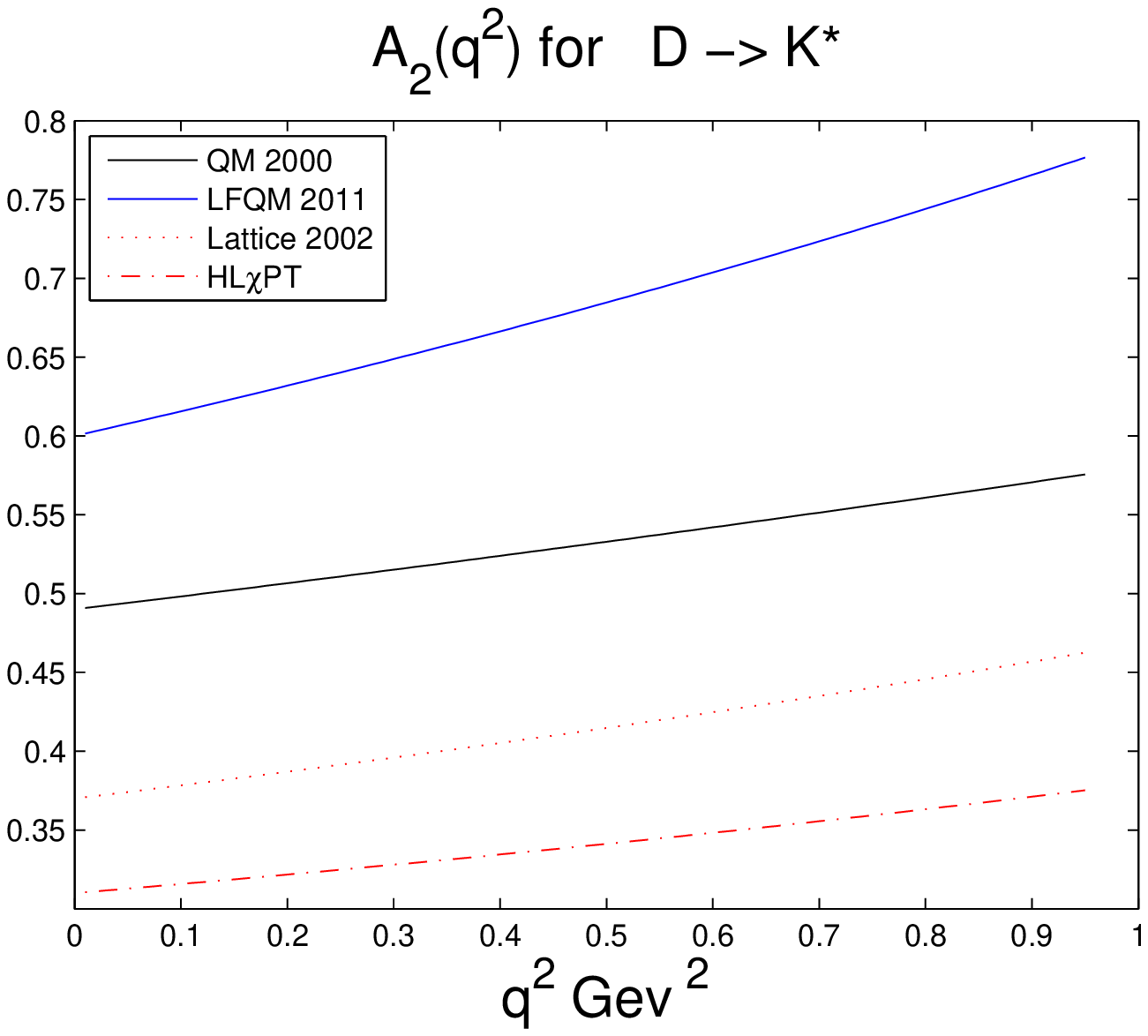}
\label{fig:dkstara2}
}
\subfigure{
\includegraphics[width=.48\textwidth]{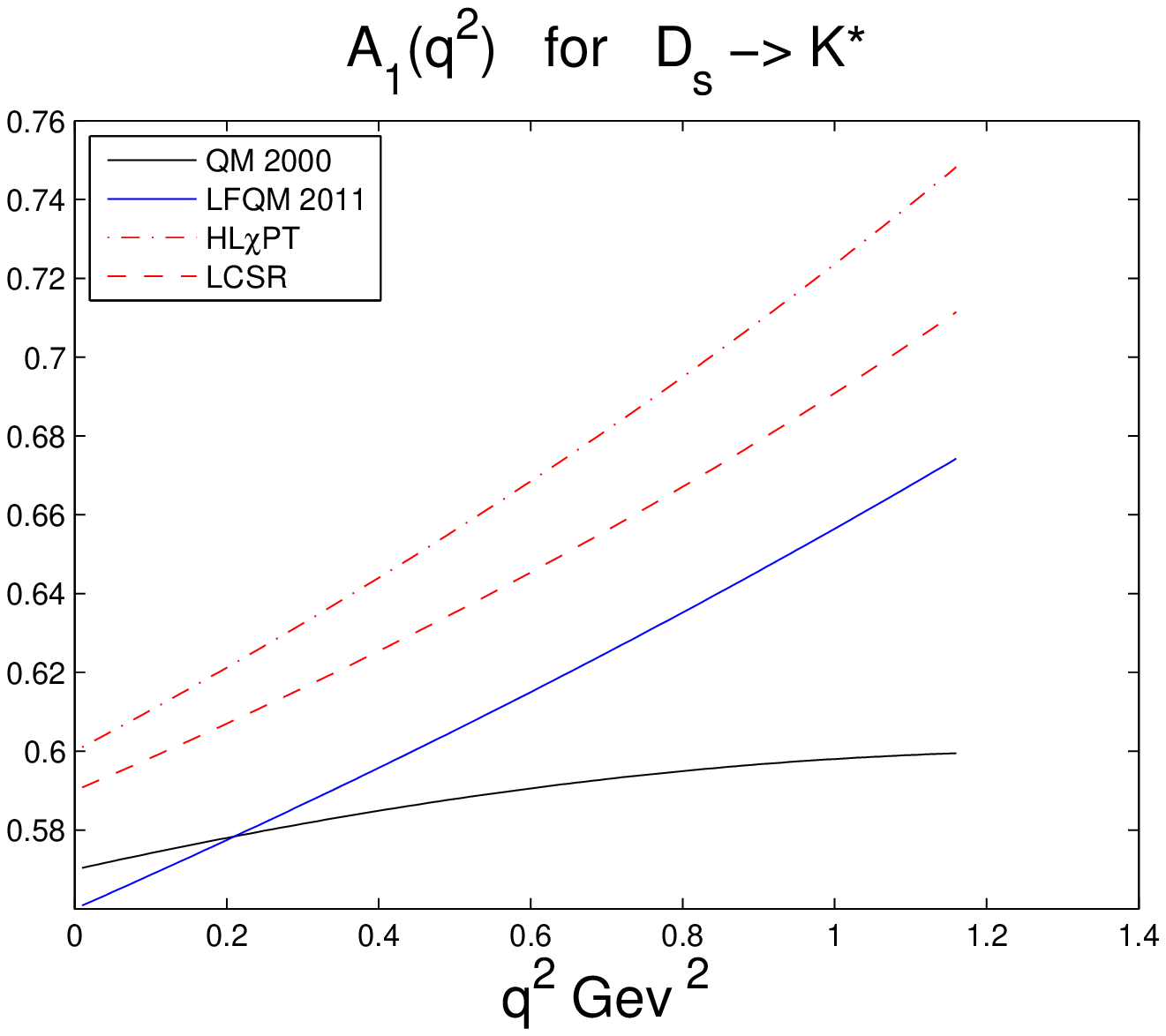}
\label{fig:dskstara1}
}
\subfigure{
\includegraphics[width=.48\textwidth]{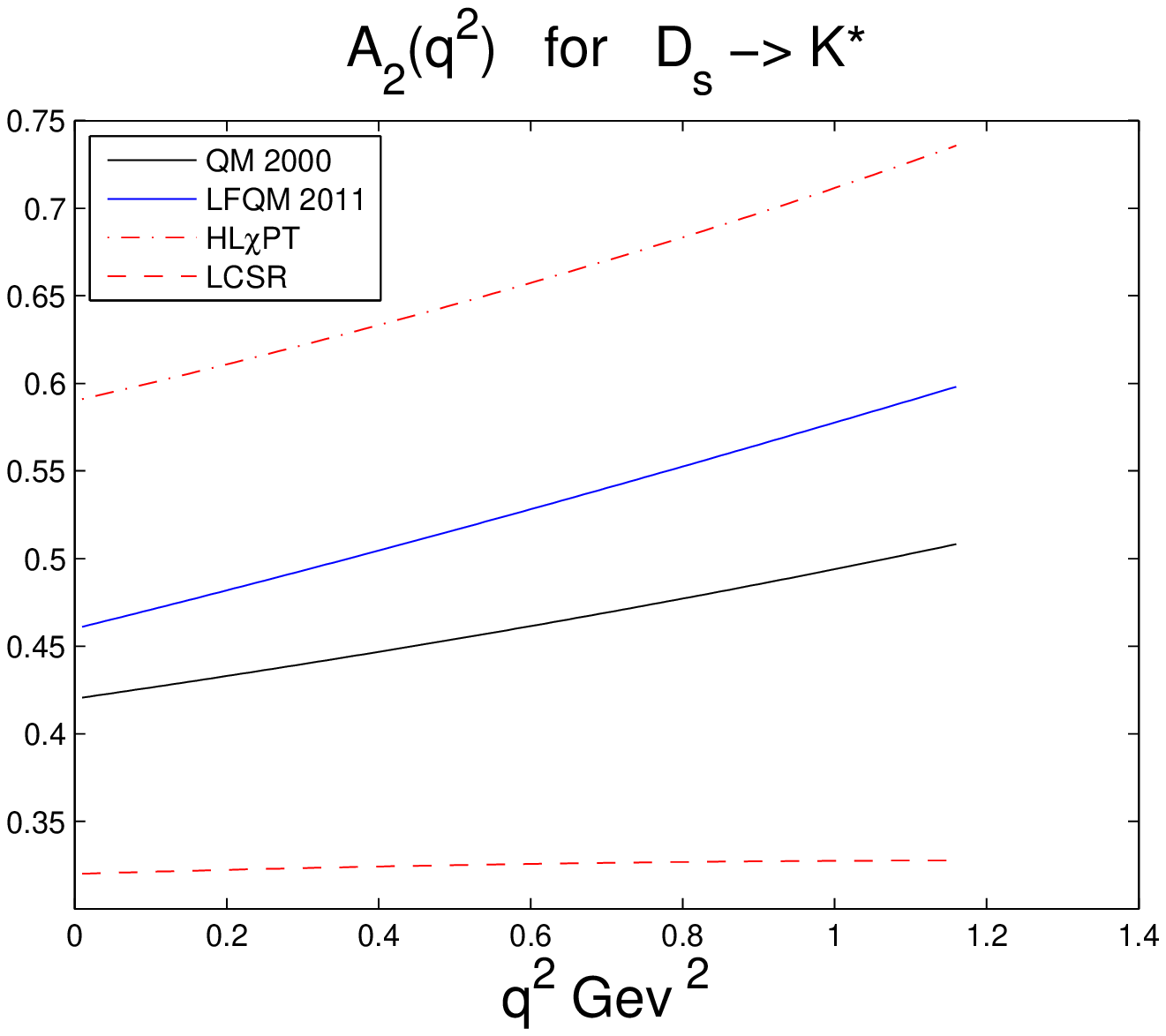}
\label{fig:dskstara2}
}
\label{fig:veca1a2}
\caption[Optional caption for list of figures]{$D \rightarrow V$
 form factors comparing frameworks used:
Data CLEO is from \cite{CLEO:2011ab}, 
 LCSR 2006 from \cite{Wu:2006rd},
LFQM  from \cite{Verma:2011yw},
Lattice 1998 from \cite{Flynn:1997ca}, Lattice 2002  from
\cite{Abada:2002ie}, HL$\chi$PT from \cite{Fajfer:2005ug}, and 
QM is from \cite{Melikhov:2000yu}.}
\end{figure}

\begin{figure}[!ht]
\centering
\subfigure{
\includegraphics[width=.48\textwidth]{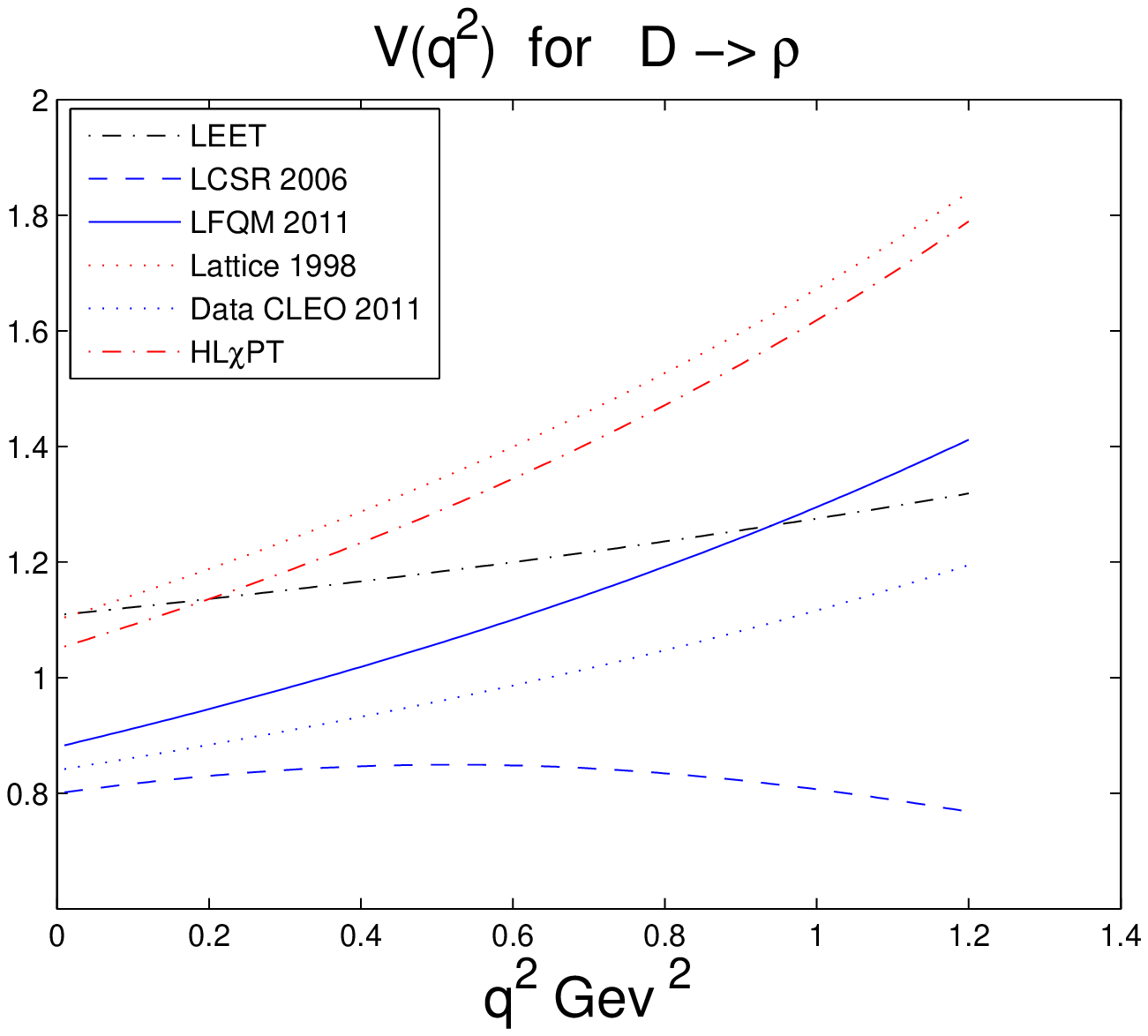}
\label{fig:drhov}
}
\subfigure{
\includegraphics[width=.48\textwidth]{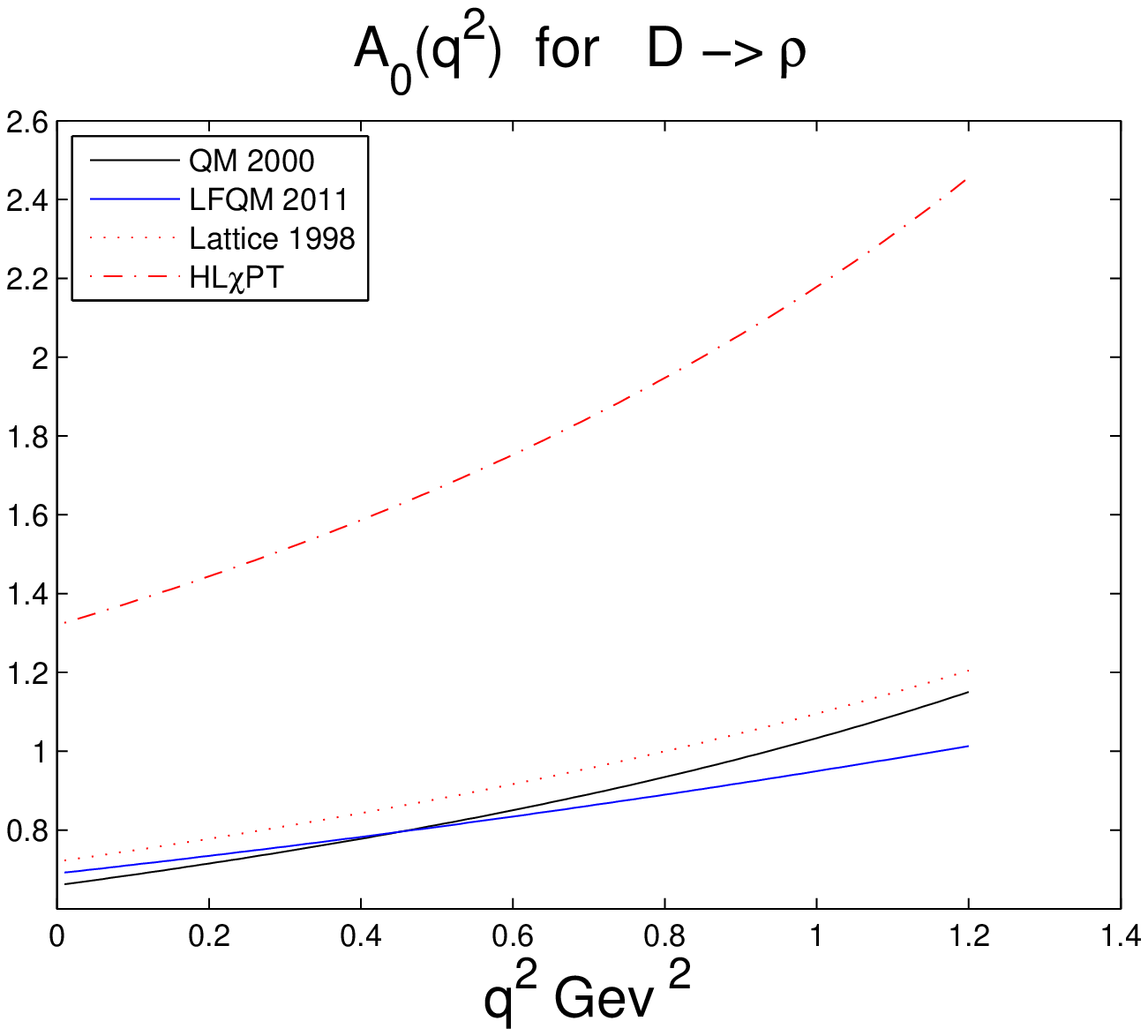}
\label{fig:drhoa}
}
\subfigure{
\includegraphics[width=.48\textwidth]{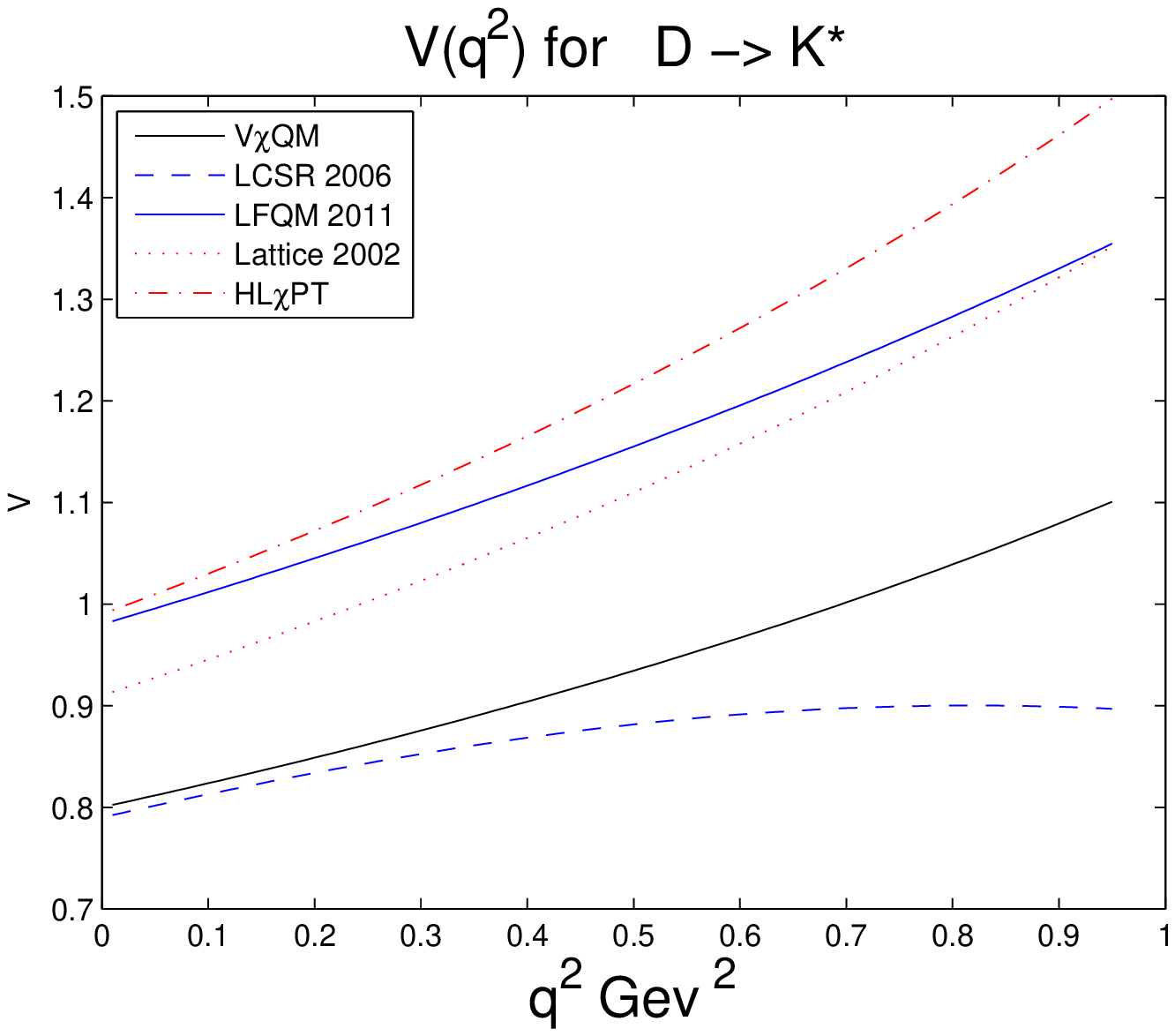}
\label{fig:dkstarv}
}
\subfigure{
\includegraphics[width=.48\textwidth]{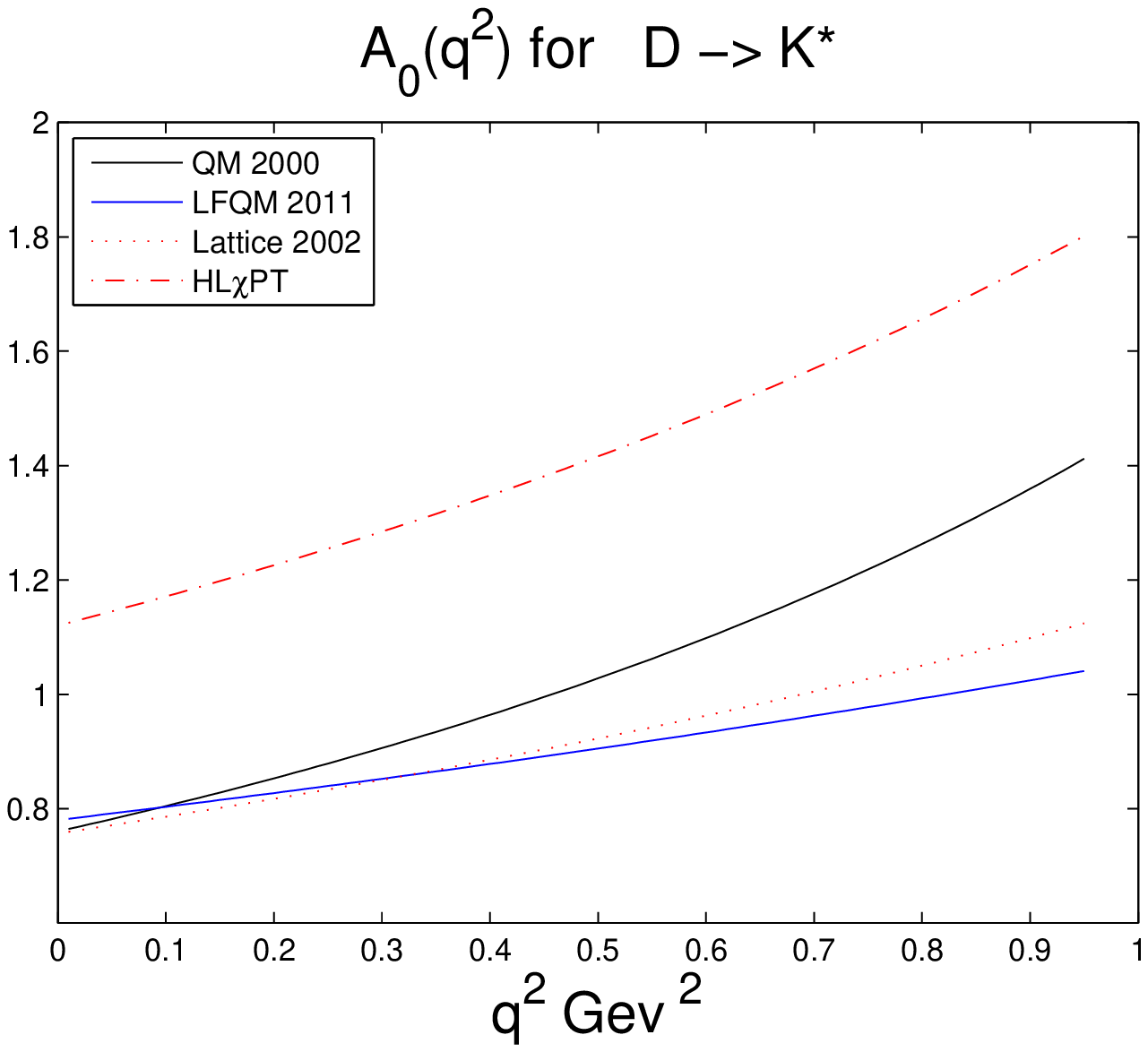}
\label{fig:dkstara}
}
\subfigure{
\includegraphics[width=.48\textwidth]{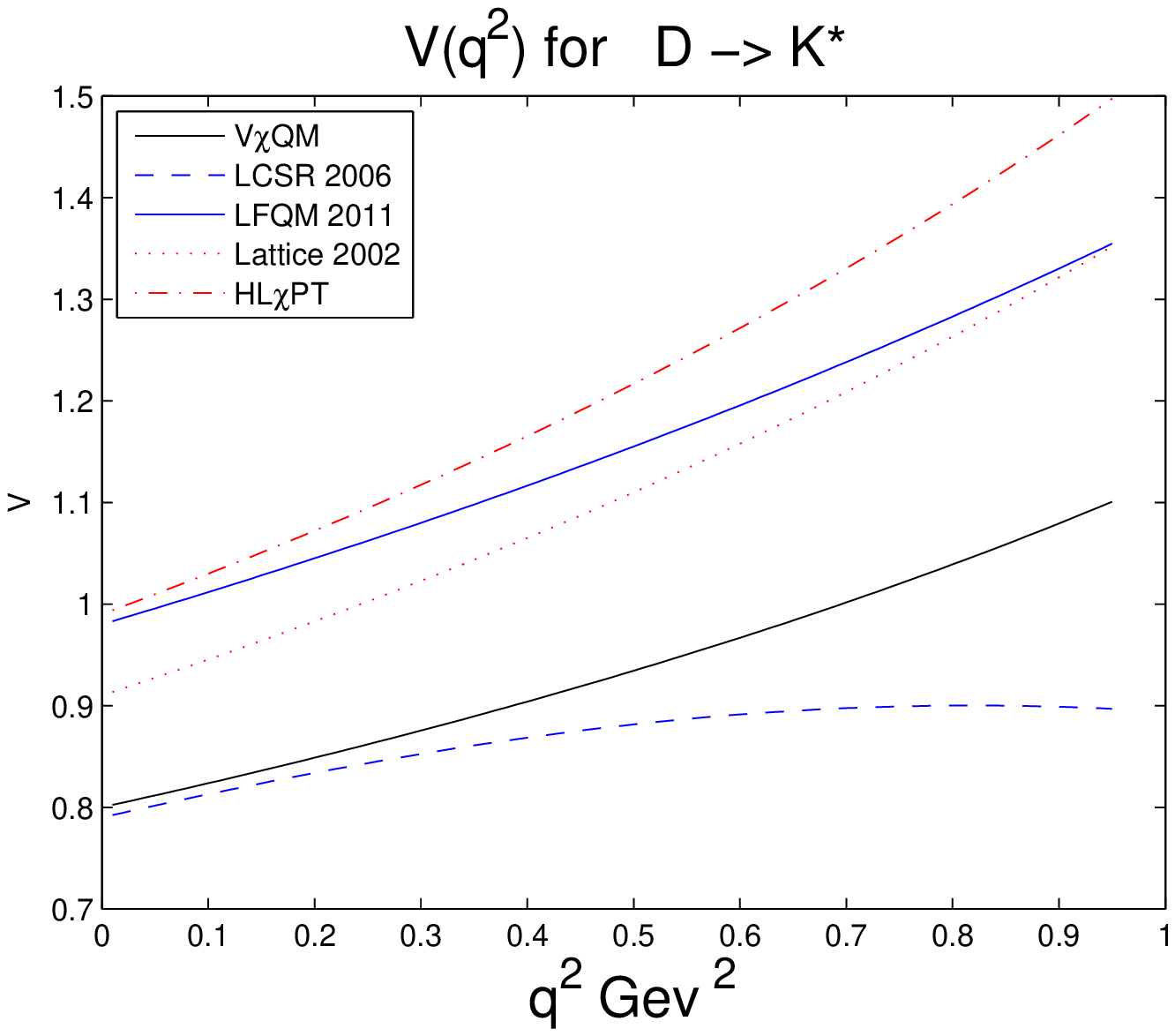}
\label{fig:dskstarv}
}
\subfigure{
\includegraphics[width=.48\textwidth]{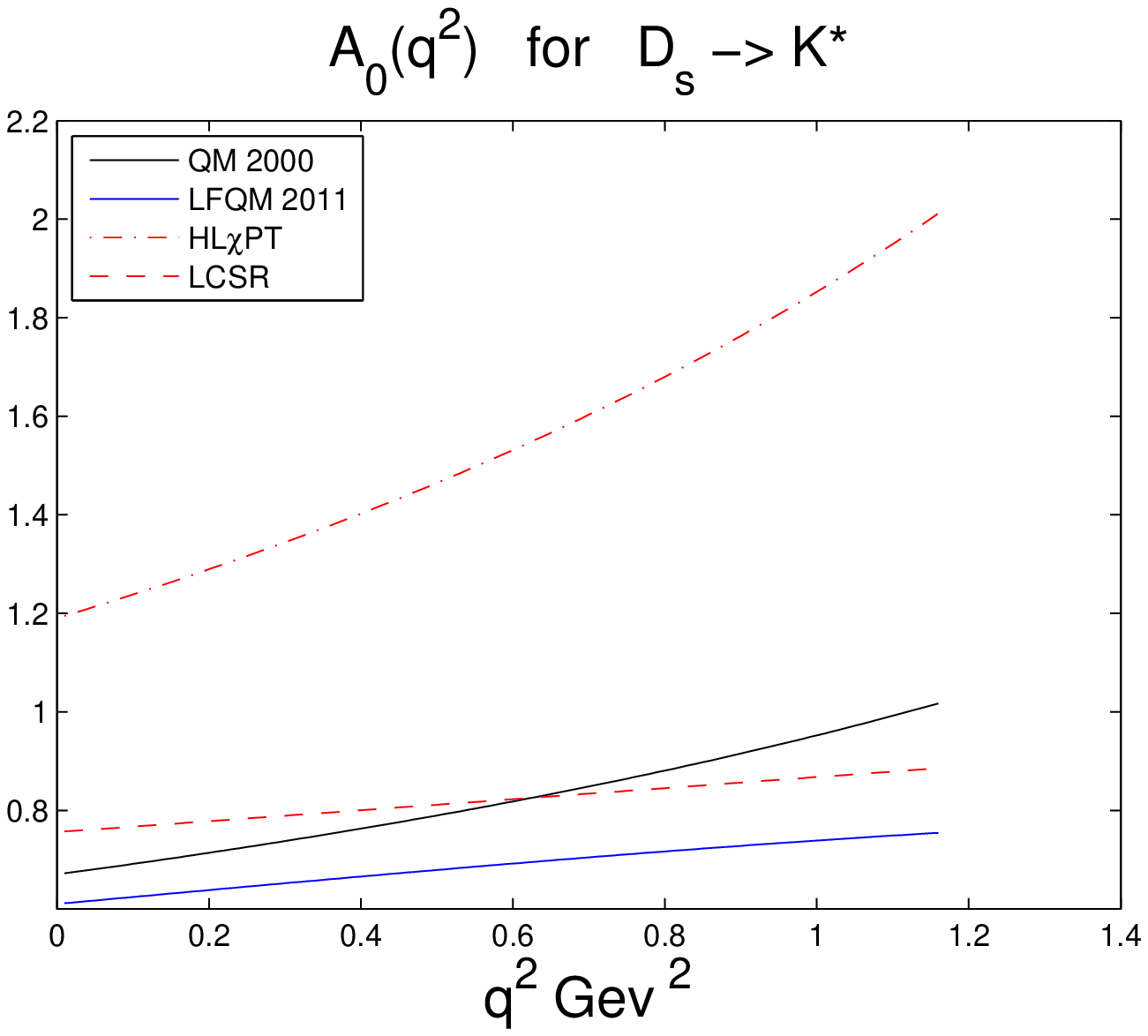}
\label{fig:dskstara}
}
\label{fig:vecffva0}
\caption[Optional caption for list of figures]{$D \rightarrow V$
 form factors comparing frameworks used: Data CLEO is from \cite{CLEO:2011ab},
 LCSR 2006 from \cite{Wu:2006rd},
LFQM  from \cite{Verma:2011yw},
Lattice 1998 from \cite{Flynn:1997ca}, Lattice 2002  from
\cite{Abada:2002ie}, HL$\chi$PT  from \cite{Fajfer:2005ug}, and 
QM is from \cite{Melikhov:2000yu}. }
\end{figure}

\begin{table}[!ht]
\begin{tabular}{l|c|c|c|c}
\hline
Decay & \,$F_+(0) $ \,& \,$F_+(0)_\chi$\, & $\zeta$ & $\zeta_1$ \\
\hline
$D \rightarrow \pi$    & \, 0.67 & \, 0.96  & \, 0.65 \, & \, 0.43 \\
$D \rightarrow K$  \,  & \, 0.74 & \, 1.06  &    0.74    &    0.49 \\
$D_s \rightarrow K$ \, & \, 0.74 & \, 1.12  &    0.74    &    0.49  \\
$D \rightarrow \eta$   & \, 0.55 & \, 0.66  &    0.55    &    0.37 \\
$D \rightarrow \eta'$  & \, 0.45 & \, 0.55  &    0.45    &    0.30  \\
\end{tabular}
\caption{Form factors for $D \rightarrow P$ at $q^2 =0$.
 The values for $F_+(0)$ are 
taken from data when
available and from sum rules for $D \to \eta,\eta'$. 
The values for $F_+(0)_\chi$
 are obtained from  \cite{Fajfer:2004mv}. A similar value (based on similar
 curve) was obtained in \cite{Hiorth:2002pp}. 
HL$\chi$QM. }
\end{table}

\begin{table}[!ht]
\begin{tabular}{l|c|c|c|c|c|c|c|c|c|c}
 Decay & \, \,$V(0)$\, & \,$V(0)_\chi$\, & \, $A_0(0)$\, & \,$A_0(0)_\chi$\, & \, $\zeta_\perp$   & \, $\zeta^{(a)}_\perp$  & \, $\zeta_\parallel$\,\, & \,  $A_1(0)$\, & \, $A_2(0)$ \\
\hline
 $D \rightarrow \rho$ \,  &  0.84  &  0.72 & 0.65 & 0.78 & \, 0.59 \, & \, 0.68 \, & \, 0.44 \, & \, 0.56 & \, 0.47 \\
 $D \rightarrow K^*$  \,  &  0.91 &  0.76 & 0.76 & 0.62 &    0.61   &   0.88    & \, 0.47 & \, 0.62 & \, 0.37 \\
 $D_s \rightarrow K^*$    & 0.77  &  0.76 & 0.76 & 0.62 &    0.53   &   0.78    & \, 0.53  & \, 0.59 & \, 0.32\\
\end{tabular}
\caption{Form factors for $D \rightarrow V$ at $q^2=0$.
 The values for $V(0)$ and $A_0(0)$ are 
taken from sum rules for  $D_s \to K^*$
and lattice calculations for $D \to K^*,\rho$. The values for
$V(0)_\chi$ and  $A_0(0)_\chi$ are obtained from
HL$\chi$QM, including vectors.}
\end{table}
From the plots we  extract  approximate  values for the
form factors $F_+(q^2)$, $V(q^2)$, $A_1(q^2)$ and $A_2(q^2)$ at $q^2 =0$.
These are collected in table I and II. But some of the data are 
uncertain, and we must expect, say, of order 20$\%$ uncertainty. 
 Using the relations  (\ref{FormfHV}) and
(\ref{zetarel}), we will obtain 
 a reasonable overall fit for the $\zeta$'s:
\begin{equation}
 \zeta \simeq 0.5  \; \,  , \;  \zeta_1  \simeq 0.3  \; \, , \;
\zeta_\perp \simeq 0.6  \; \, , \;  \zeta_{||}  \simeq 0.5 \; \, , 
\; \zeta_\perp^{(a)} \simeq 0.7 \; \, .   
\label{zetas}
\end{equation}

\section{Conclusions}

We have collected present information on various form factors
for the transitions $D \rightarrow P$ and $D \rightarrow V$ 
($P$= pseudoscalar, $V$= vector)
obtained from
various methods and sources like data, lattice gauge theory, LCSR , etc. 
From the curves we have as far as possible determined the values of
 relevant form factors at $q^2 = 0$, and then extracted values
 for the LEET form factors $\zeta_i$.
Our LE$\chi$QM gives relations between the $\zeta_i$'s.
We have previously found \cite{Leganger:2010wu} $\zeta_1/\zeta \sim m/E$.
Here we have in addition shown  that $\zeta_\perp^{(a)} \rightarrow \zeta_\perp$
 for $m/E \rightarrow 0$ as it should.

The values of $V(0)$ from the plots show a large variation among
 the various methods used,
which makes $\zeta_\perp$ uncertain. However   $\zeta_\perp$ is also related to 
$A_0(0)$ such that we obtain a reasonable fit in eq. (\ref{zetas}).
We observe what we expected, namely 
that LEET works best for $q^2$
 close to zero, while HL$\chi$QM (eventually supplemented by HL$\chi$QM)
 works best close to the no-recoil point.

The LEET form factors  $\zeta$ and $\zeta_\perp$
 will  determine the coupling constants 
$G_A$ and $G_V$, 
 which may be used in calculation of
nonfactorizable (color suppressed)
 non-leptonic $D$-meson decays, in the same manner as have previously
 been done for
$K \rightarrow \pi \pi$ \cite{Bertolini:1997nf,Bertolini:2000dy},
  $D \rightarrow K^0 \overline{K^0}$ \cite{Eeg:2001un}, 
$B \rightarrow D \overline{D}$ \cite{Eeg:2003xe,Eeg:2005au},
 $B \rightarrow D \pi$ \cite{Leganger:2010wu}, and 
$B \rightarrow \pi^0 \pi^0$ \cite{Eeg:2010rk}.
 Then non-leptonic decay amplitudes might be written in terms of the
 LEET form factors  $\zeta_i$, both for the factorized and the color 
suppressed cases.

\bibliography{particle}

\end{document}